\documentclass[numberwithinsect]{lipics-v2021}

\usepackage{mathrsfs}
\usepackage{aligned-overset}    %
\usepackage{tikz}
\usetikzlibrary{cd,decorations.pathmorphing}
\tikzset{>=to}
\tikzcdset{arrow style=tikz}

\makeatletter

\DeclareFontFamily{U}{hira}{}
\DeclareFontShape{U}{hira}{m}{n}{<-> udmj30}{}
\newcommand\yo{{\!\text{\usefont{U}{hira}{m}{n}\symbol{'207}}\!}}

\newtheorem{question}[theorem]{Question}

\let\paragraph\subparagraph

\let\cat\mathcal
\let\equals\approx      %
\let\cong\simeq         %
\let\mkPrsh\vartriangle %
\let\mkStr\triangledown %

\newcommand{\ARG}{\mathord{-}}

\newcommand{\NP}{\textsf{NP}}
\newcommand{\Ptime}{\textsf{P}}

\DeclareMathOperator{\plainCSP}{CSP}
\DeclareMathOperator{\hCSP}{CSP_\textit{hom}}
\DeclareMathOperator{\hPCSP}{PCSP_\textit{hom}}
\DeclareMathOperator{\sCSP}{CSP_\textit{sat}}
\DeclareMathOperator{\sPCSP}{PCSP_\textit{sat}}
\DeclareMathOperator{\ppCSP}{CSP_\textit{pp}}
\DeclareMathOperator{\ppPCSP}{PCSP_\textit{pp}}
\newcommand{\PLC}[1]{\sPCSP(\id, {#1})}
\DeclareMathOperator{\PMC}{PMC}

\DeclareMathOperator{\id}{id}
\DeclareMathOperator{\pol}{pol}
\newcommand\freeStr[2]{\mathbf F_{#2}(#1)}
\NewDocumentCommand{\Lan}{ e{_} }{
  \operatorname{Lan} \IfValueT{#1}{_{#1}\!}}
\NewDocumentCommand{\Ran}{ e{_} }{
  \operatorname{Ran} \IfValueT{#1}{_{#1}\!}}
\DeclareMathOperator*{\colim}{colim}

\DeclareMathOperator{\gr}{gr}
\newcommand{\el}{\int\!}
\newcommand{\gl}[1]{\langle #1 \rangle}

\newcommand{\nerve}[1]{\mathsf N_{#1}}
\newcommand{\kay}[1]{\widehat{#1}}
\newcommand{\kayyo}{\mathord{%
  \mathrlap{\raisebox{-3pt}{$\widehat{\phantom\yo}$}}\yo}}

\DeclareMathOperator{\@p}{op} \newcommand\op{{\@p}}
\newcommand\catfont{\normalfont\sffamily}
\newcommand{\Str}{{\text{\catfont Str}}}
\newcommand{\graphs}{{\text{\catfont Gra}}}
\newcommand{\set}{{\text{\catfont Set}}}
\newcommand{\Cat}{{\text{\catfont Cat}}}
\newcommand{\fcat}{{\text{\catfont FCat}}}
\newcommand{\fin}{{\text{\catfont Fin}}}
\newcommand{\str}[1]{\Str_{#1}}
\newcommand{\comma}{\mathbin{\downarrow}}

\newcommand{\psh}[1]{[#1, \fin]}
\newcommand{\pshcat}[1]{\psh{\cat #1}}

\numberwithin{equation}{section}

\newcommand{\Yes}{\textsc{Yes}}
\newcommand{\No}{\textsc{No}}

\makeatother
 \nolinenumbers

\title{A categorical perspective on constraint satisfaction:
       The wonderland of adjunctions}

\author{Maximilian Hadek}
        {Charles University, Prague, Czech Republic}
        {maximilian.hadek@matfyz.cuni.cz}
        {https://orcid.org/0009-0008-2102-7567}
        {Funded by the European Union (ERC Synergy Grant No.101071674,
         POCOCOP).
         Views and opinions expressed are however those of the
         authors only and do not necessarily reflect those of the European
         Union or the European Research Council Executive Agency. Neither the
         European Union nor the granting authority can be held responsible for
         them.}

\author{Tom\'a\v{s} Jakl}
        {Czech Technical University, Prague, Czech Republic}
        {tomas.jakl@cvut.cz}
        {https://orcid.org/0000-0003-1930-4904}
        {Partially funded by the EXPRO project 20-31529X awarded by the Czech
         Science Foundation, and supported by the Institute of Mathematics,
         Czech Academy of Sciences (RVO 67985840). Also funded by the European
         Union’s Horizon 2020 research and innovation programme under the Marie
         Skłodowska-Curie grant agreement No~101111373.}

\author{Jakub Opr\v{s}al}
        {University of Birmingham, Birmingham, UK}
        {j.oprsal@bham.ac.uk}
        {https://orcid.org/0000-0003-1245-3456}
        {}

\titlerunning{A categorical perspective on constraint satisfaction}
\authorrunning{M. Hadek, T. Jakl, and J. Opršal}

\Copyright{Maximilian Hadek, Tomáš Jakl, and Jakub Opršal}

\funding{}



\relatedversiondetails[cite=HadekJO26]{Conference version}
  {https://doi.org/10.4230/LIPIcs.LICS.2026.22}
\hideLIPIcs

\acknowledgements{In the past, several people have thought about different
  angles of the categorical perspective on the constraint satisfaction problem.
  We would like to thank Marcin Wrochna, who has kindly shared his categorical
  notes on CSPs with us, and Libor Barto for the insights he provided to us
  over many years.
  We would like to thank Anuj Dawar for providing funding for Jakub Opršal's
  visit to Cambridge in 2022, which initiated the work.}

\begin{document}

\maketitle

\begin{abstract}
  The so-called algebraic approach to the constraint satisfaction problem (CSP)
  has been a prevalent method of the study of complexity of these problems
  since early 2000's. The core of this approach is the notion of
  \emph{polymorphisms} which determine the complexity of the problem (up to
  log-space reductions).  In the past few years, a new, more general version of
  the CSP emerged, the promise constraint satisfaction problem (PCSP), and the
  notion of polymorphisms and most of the core theses of the algebraic approach
  were generalised to the promise setting. Nevertheless, recent work also
  suggests that insights from other fields are immensely useful in the study of
  PCSPs including algebraic topology.

  In this paper, we provide an entry point for category-theorists into the
  study of complexity of CSPs and PCSPs. We show that many standard CSP notions
  have clear and well-known categorical counterparts. For example, the
  algebraic structure of polymorphisms can be described as a set-functor
  defined as a right Kan extension.  We provide purely categorical proofs of
  core results of the algebraic approach including a proof that the complexity
  only depends on the polymorphisms.  Our new proofs are substantially shorter
  and, from the categorical perspective, cleaner than previous proofs of the
  same results. Moreover, as expected, they are applicable more widely.  We
  believe that, in particular in the case of PCSPs, category theory brings
  insights that can help solve some of the current challenges of the field.
\end{abstract}

\section{Introduction}

If $\Ptime\neq\NP$, what distinguishes problems in $\Ptime$ from those that are $\NP$-complete?
This question is one of the key motivations for the systematic study of the complexity of \emph{constraint satisfaction problems (CSPs)}.
The goal of a CSP is, given an input instance consisting of variables attaining values in some domain and a list of constraints, each involving a bounded number of variables, to decide whether there exists an assignment of values to variables that satisfies all constraints simultaneously.
By varying which constraints are allowed to appear in input instances, we obtain many well-known combinatorial problems including Boolean satisfiability (SAT) and its variants, e.g.\ HornSAT and 2SAT, graph $k$-colouring, systems of linear equations over a fixed ring, linear programming, and many others. Formally, the restriction of constraints is given as a single relational structure $A$ called a~\emph{template}.

Considerable effort has been dedicated to the complexity classifications of \emph{finite-template CSPs} (i.e.\ where the template is a finite relational structure). The most successful theory in addressing the complexity of these problems is the so-called \emph{algebraic approach to CSPs}.
This theory started with the work of Jeavons \cite{jeavons1998algebraic}, and has been further developed over the past three decades, including the work of Bulatov, Jeavons, and Krokhin \cite{BulatovJK05}, which gave the theory its current name.
The core idea of this theory is to assign to each template $A$ an abstract algebraic structure, denoted by $\pol(A)$, which determines the complexity of the problem up to log-space reductions.
A major achievement of this theory is the \emph{CSP Dichotomy Theorem} which states that, for each finite structure $A$, either $\plainCSP(A)$ is $\NP$-complete or it is in $\Ptime$. The statement was first conjectured by Feder and Vardi \cite{federvardi1998} and independently confirmed by Bulatov \cite{Bulatov17} and Zhuk \cite{Zhuk20}.
More importantly, the line between \NP-complete problems and problems in \Ptime{} is precisely described by the algebraic invariants $\pol(A)$ of the problem, and hence the dichotomy provides a somewhat satisfactory answer to our opening question.

In this paper, we revisit the general theory of the algebraic approach through more abstract lenses of category theory; we have two key motivations for this.
Firstly, categorical perspectives seem to appear (with or without the knowledge of the authors) more-and-more often in recent work on the complexity of CSPs, and in particular, in the study of the complexity of \emph{promise constraint satisfaction problems (PCSPs)}. The latter is a natural generalisation of CSPs from decision to promise problems including, e.g.\ the problem of distinguishing graphs that are $3$-colourable from those that are not even $6$-colourable.
Examples of the usage of a categorical language in the PCSP literature include the definition of an \emph{abstract minion} given by Brakensiek, Guruswami, Wrochna and Živný \cite{BrakensiekGWZ20}, which coincides with the notion of a \emph{set functor}, or efficient reductions between PCSPs in the form of \emph{adjoint functors} used by Wrochna and Živný \cite[Observations 2.1 and 2.2]{WrochnaZ20}.
Secondly, a recent direction in the study of PCSPs is based on applying methods from algebraic topology, starting with Krokhin, Opršal, Wrochna, and Živný \cite{KOWZ23} and following with Filakovský, Nakajima, Opršal, Tasinato, and Wagner~\cite{FilakovskyNOTW24}, Avvakumov, Filakovský, Opršal, Tasinato, and Wagner~\cite{AvvakumovFOTW25}, and by Meyer and Opršal \cite{MeyerO25}.
Since algebraic topology has intimate connections to category theory, a lot of these results already have a clear category-theoretic flavour.
Finally, let us also mention that Ó~Conghaile \cite{OConghaile22} and Abramsky \cite{Abramsky22} provided a sheaf-theoretic perspective on \emph{$k$-consistency} and similar algorithms.

Although categorical thinking is already used in the CSP research, and has led to several key advances, e.g.\ \cite{BBKO21,KOWZ23}, most authors do not state or explain their results in the categorical language.
We believe that a proper categorical exposition is long overdue. This paper is the first step in that direction.

\begin{figure*}[ht]
\centering
\bgroup
\def\arraystretch{1.1}
\def\TODO{{\color{blue}TODO}}
\begin{tabular}{|l|l|l|}
    \hline
    \bf CSP notion (used in \cite{BBKO21}) & \bf CT notion & \bf Reference \\
    \hline
    minion & functor \(\fin \to \fin\) & \cite{BrakensiekGWZ20} \\
    minion homomorphism & natural transformation &  \\
    first-order structure & functor \(\cat S \to \fin\) & \cite[Proposition 3.2.5]{PareMakkai90} \\
    \hline
    gadget replacement & Yoneda extension & Section~\ref{s:gadget-reductions} \\
    pp-interpretation  & nerve & Appendix~\ref{sec:pp-interpretations} \\ %
    \hline
    \(\pol(A,B)\) & \(\Ran_A B\) & Lemma~\ref{lem:pol-is-ran} \\
    free structure \(\freeStr A M\) & \(M \circ A\) & Example~\ref{ex:free-structure} \\
    \hline
    minor condition \(\Sigma\) & functor \(D \colon \cat J\to \fin\)  &  Section~\ref{sec:minor-conditions} \\
    & \hfill (for a finite category \(\cat J\)) & \\[0.2em]
    \(M \models \Sigma\) &  \(\lim(M\circ D) \not= \emptyset\) &  \\
    the indicator structure \(\mathbf I_{\,\Sigma}(A)\) & \(\gl D \circ A\) & \\
    the minor condition \(\Sigma(A, B)\)  & \(A \circ \gr B\) &  \\
    \hline
\end{tabular}
\egroup
\caption{Translation table of common constraint satisfaction notions and their category-theoretic counterparts.}
\label{fig:dictionary}
\end{figure*}
 
\paragraph*{Our contributions}

The core contribution of this paper is conceptual. We build a bridge between the algebraic CSP and categorical model theory communities by providing a translation between combinatorial constructions used in the CSP theory and standard notions in categorical model theory; see table in Fig.~\ref{fig:dictionary} for an overview. This connection allows us to simplify (from the categorical perspective) the proofs of the core theorems of the algebraic approach to CSPs in their latest refinement \cite{BartoOP18,BBKO21}; our proofs follow the exposition in \cite{BBKO21} which generalises \cite{BartoOP18} to PCSPs providing new proofs for non-promise setting as well.

The main difference from the earlier algebraic treatment \cite{BBKO21} is the introduction of a new equivalent definition of finite-template CSP (Definition~\ref{def:pcsp-v2}), which despite the fact that most of Section~\ref{sec:two-pcsps} is folklore in some communities, has not appeared before. Using this definition, we provide two separate proofs (one in Section~\ref{sec:square} and one in Section~\ref{sec:gadgets}) of core theorems of the algebraic approach.

As it often happens when categorising existing results, our statements are more general than the original proofs. In particular, our theorems are applicable to multisorted relational structures, used recently by Dalmau and Opršal \cite{DalmauO24} and by Barto, Butti, Kazda, Viola, and Živný \cite{BartoButti+24}, relational structures with unary operations, locally finite simplicial sets, etc.
Furthermore, we give a succinct proof of the hardness part of the Bulatov--Zhuk dichotomy, which was originally proved by Bulatov, Jeavons, and Krokhin \cite{BulatovJK05}. We also generalise the Bulatov--Zhuk dichotomy to multisorted structures with relational symbols and unary function symbols.
Our exposition also opens up the possibility of further generalisations and exploration of CSPs of more varied structures including algebraic structures, ordered structures, simplicial sets, or combinations of thereof.

\subsection{Our approach}

As a starting point of our purely categorical formulation, we look at the finite-template CSP problem from two different angles.
Firstly, we treat it as a homomorphism problem of finite copresheaves.
Secondly, we start with a natural categorical formulation of the \emph{uniform CSP} where all constraints are allowed, and explain restrictions that make the problem computationally easier.

\paragraph{Fixed-template CSPs}

Traditionally, \emph{fixed-template CSPs} are defined as homomorphism problems for finite relational structures. Instead, we treat these problems as homomorphism problems of \emph{finite copresheaves}, i.e.\ functors of the type $\cat S \to \fin$ where $\cat S$ is a finite category (with finitely many objects and finitely many morphisms) and~\(\fin\) is the category of finite sets and functions.
A finite relational structure can be treated as such a copresheaf (see Section~\ref{sec:relational-structures}) hence this scope generalises relational structures.

\begin{definition}[Finite-template CSP] \label{def:csp-v1}
  Let $A\colon \cat S \to \fin$ be a copresheaf on a finite category $\cat S$. We define $\hCSP(A)$ as the following problem: Given another copresheaf $X \colon \cat S \to \fin$, decide whether there is a natural transformation $h\colon X \to A$.
\end{definition}

The core of the algebraic approach is a sufficient criterion for the existence of log-space reductions between such CSPs, which we phrase using \emph{right Kan extensions}, denoted by $\Ran$.

\begin{theorem} \label{thm:1}
  Let $A \colon \cat S \to \fin$ and $B \colon \cat T \to \fin$ be two copresheaves on finite categories $\cat S$ and $\cat T$.
  If there is a natural transformation $\Ran_B B \to \Ran_A A$, then $\hCSP(A)$ reduces to $\hCSP(B)$ in log-space.
\end{theorem}

In the usual algebraic statement of the above theorem, see e.g.\ \cite[Theorem 1.3]{BartoOP18}, $\pol(A)$ would be used in place of $\Ran_A A$. In fact, we show that $\pol(A)$, interpreted as a functor $\fin\to \fin$, is an explicit description of $\Ran_A A$ (see Lemma~\ref{lem:pol-is-ran}).
Note that $\Ran_A A$ is the codensity monad of \(A\), nevertheless the additional monad structure is irrelevant in the above theorem, which becomes apparent in the generalisation of the above theorem to PCSPs (see Theorem~\ref{thm:gadgets}).

In light of the above theorem, the Bulatov--Zhuk dichotomy theorem looks as follows.

\begin{theorem}[Bulatov--Zhuk] \label{thm:bulatov-zhuk-for-presheaves}
  Let $A\colon \cat S \to \fin$ be a copresheaf on a finite category~$\cat S$. Then we have the following dichotomy:
  \begin{enumerate}
    \item If there is a natural transformation $\Ran_A A \to \id_\fin$, then $\hCSP(A)$ is \NP-complete.
    \item If there is no such natural transformation, then $\hCSP(A)$ is in \Ptime.
  \end{enumerate}
\end{theorem}

Our proof of this theorem leans on the original theorem of Bulatov--Zhuk \cite{Bulatov17,Zhuk20}, by employing a translation from copresheaves to relational structures which preserves the structure of $\pol(A)$.
Independently of this, we provide a categorical proof of the hardness in case (1), cf.\ Corollary~\ref{cor:bulatov-zhuk-NP-complete}.

\paragraph{The uniform CSP and its promise version}

In plain words, the goal of the uniform CSP is to assign values to variables subject to some constraints, hence an (input) instance of the CSP is a finite list of variables $I$ where each variable $i \in I$ is given a set $D_i$ of allowed values.
The constraints are given by relations, i.e.\ each constraint is an expression of the form $(i_1, \dots, i_k) \in R$ for some $R \subseteq D_{i_1} \times \dots \times D_{i_k}$.
In this paper, we use a slightly different, yet equivalent, reformulation which allows for an easier categorical treatment.

\begin{definition}[Uniform CSP] \label{def:uniform-csp}
  We define the \emph{uniform constraint satisfaction problem} as follows:
  An input is a diagram \mbox{$D \colon \cat J \to \fin$} where $\cat J$ is a finite category whose objects and morphisms we call \emph{variables} and \emph{constraints}, respectively.
  The task is to decide whether $D$ admits a \emph{solution}, that is, there is an assignment $d_i \in D(i)$ for each \emph{variable}~$i$ such that, for every \emph{constraint} $f\colon i\to j$,\,  \mbox{$D(f)(d_i) = d_j$}.
\end{definition}

Categorically speaking, the goal of the uniform CSP is to decide whether the limit of the diagram $D$ is non-empty.
The (uniform) CSP is a fundamental NP-complete problem -- it directly encodes Boolean SAT, graph 3-colouring, and many other well-known combinatorial problems.
Therefore, some refinements of the formulation are required to make the problem possibly solvable in polynomial-time. There are two natural ways to simplify the problem: restricting what constraints are allowed or allowing approximate solutions.

\subparagraph{Restricting the allowed constraints}

A natural way to restrict the allowed constraints is to require that an instance $D \colon \cat J \to \fin$ factors through a given functor $A \colon \cat S \to \fin$. The functor $A$ plays the role of the template here and, as before, we require that $A$ is a finite template, i.e.\ \(\cat S\) is a finite category.

\begin{definition}[Finite-template CSP, version 2] \label{def:csp-v2}
  Fix a copresheaf $A \colon \cat S \to \fin$ on a finite category $\cat S$. We define $\sCSP(A)$ as follows: The input is a functor $D \colon \cat J \to \cat S$ where $\cat J$ is a finite category, and the goal is to decide whether the functor $A\circ D\colon \cat J \to \fin$ has a solution (i.e.\ the limit of $A\circ D$ is non-empty).
\end{definition}

In Theorem~\ref{thm:hom=fib} we show that this version of finite-template CSP is equivalent to the homomorphism problem for $A$ (from Definition~\ref{def:csp-v1}).
The proof employs the \emph{(discrete) Grothendieck construction} and its left adjoint for the two directions of the reductions.
Naturally, we argue that both of these are efficiently computable.

\subparagraph{Approximate solutions}

The search for efficient algorithms that provide approximate solutions to NP-complete problems is an active field of algorithmic research.
In our case, we deal with `structural approximations' rather than the more common analytic version, which aims to satisfy as many constraints as possible.
Below we define a version of the uniform CSP which is a promise problem, i.e.\ the sets of \Yes- and \No-instances are disjoint but not necessarily complementary. The word `promise' indicates that the input instance is guaranteed to fall into one of these sets.

\begin{definition}[Promise version of the uniform CSP] \label{def:plc}
  Fix a functor $M \colon \fin \to \fin$ such that $M(X) \neq \emptyset$ for all non-empty $X$. We define a promise problem, which we denote $\PLC M$,\footnote{This notation is explained in Definition~\ref{def:pcsp-v2}.} as follows. Given a finite copresheaf $D\colon \cat J \to \fin$ output:
  \begin{description}
    \item[\Yes] if $D$ admits a solution; or
    \item[\No] if $M\circ D\colon \cat J \to \fin$ does not admit a solution.
  \end{description}
\end{definition}

The assumption that $M(X)\neq \emptyset$ for a non-empty $X$ is required for the disjointness of the \Yes- and \No-instances.
Note some similarities with Definition~\ref{def:csp-v2}.
In both cases, we are asked to decide the existence of a solution of a post-composition with another functor.
The type of functor we compose is different: $\cat S \to \fin$ for some finite category $\cat S$ vs.\ $\fin \to \fin$.

The second fundamental theorem of the algebraic approach to CSPs, originating in \cite[Theorem 3.12]{BBKO21}, states that every $\hCSP(A)$ is equivalent to $\PLC M$ for a suitable $M$, namely $M = \Ran_A A$. 

\begin{theorem}[informally, cf.\ Theorem~\ref{t:fundamental-theorem}] \label{thm:2}
  Let $A \colon \cat S \to \fin$ be a finite copresheaf. The problems $\hCSP(A)$ and $\PLC{\Ran_A A}$ are equivalent.
\end{theorem}

Importantly, the theorem provides a philosophical shift in the interpretation of the algebraic approach.
The philosophical consequence of Theorem~\ref{thm:1} is that, for every CSP with template $A$, we can assign algebraic invariants $\pol(A) = \Ran_A A$ that capture the complexity of the problem up to log-space reductions.
The new interpretation following Theorem~\ref{thm:2} is that, since every (promise) CSP is equivalent to a problem of the form $\PLC M$, it is sufficient to work only with the algebraic structure of \(\pol(A)\).

We prove all theorems in a strictly more general setting of \emph{promise CSPs} (see Definitions~\ref{def:csp-v1} and \ref{def:csp-v2}) which forces us to work with the right definitions and, as a result, simplify many proofs.

\subsection{Organisation of the paper}

After short preliminaries (Section~\ref{sec:preliminaries}), we present two definitions of PCSPs and their equivalence (Section~\ref{sec:two-pcsps}).
The following two sections (Section~\ref{sec:square} and \ref{sec:gadgets}) present two different proofs of the fundamental theorem of the algebraic approach to PCSPs.
In Section~\ref{sec:bulatov-zhuk}, we present the proof of Theorem~\ref{thm:bulatov-zhuk-for-presheaves}.
Finally, in Section~\ref{sec:conclusion}, we outline some directions for future research.

We present the results in categorical language and, when appropriate, follow this explanation with a comparison with the combinatorial/algebraic language commonly used to study CSPs and PCSPs.
By doing so, we hope that this paper can serve as an entry point for category-theorists into the study of complexity of CSPs and PCSPs. We also hope that CSP experts, with basic knowledge of categories, will appreciate seeing their favourite theorems from a new angle.

\section{Preliminaries}
  \label{sec:preliminaries}

We outline a few necessary concepts from category theory and the theory of PCSPs. To make the material accessible to a wide audience, we assume basic familiarity with only elementary notions of category theory, such as functors, natural transformations, adjunctions, limits, colimits as found in, e.g.~\cite{abramskytzevelekos2010introduction,Riehl2016context}.
Also, we sketch proofs of some key facts, to emphasise the accessibility of the category theory we use.

In this paper, we work mostly with finite structures, finite sets, and finite (co)limit diagrams. We denote the category of finite sets and functions between them by $\fin$.
In the rare case when we need to refer to the category of arbitrary sets, we use the notation \(\set\).
Further, we denote the category of functors from a category $\cat A$ to a category $\cat B$ by $[\cat A, \cat B]$. Morphisms in this category are natural transformations.
This paper mainly concerns categories of \emph{finite copresheaves}, i.e.\ the categories $\pshcat S$ where $\cat S$ is a finite category.
We write $A \to B$, where $A, B$ are objects of some category $\cat C$, if there exists a morphism in $\cat C$ from $A$ to $B$, and, in particular, we use a simple arrow to denote the existence of a natural transformation between functors (or copresheaves).

\subsection{Graphs and relational structures}
  \label{sec:relational-structures}

To compare the theory explored in this paper with the more traditional combinatorial approach, we give an example of how relational structures can be treated as copresheaves. This follows a well-known encoding of first-order structures as sketches, see, e.g., \cite[proof of Proposition 3.2.5]{PareMakkai90}.

\begin{example}[Digraphs as copresheaves] \label{ex:graphs-as-presheaves}
  We start with an example of \emph{simple digraphs}, i.e.\ oriented graphs without parallel edges. Combinatorially, a \emph{finite digraph} is a pair $\mathbf G = (V^\mathbf G, E^\mathbf G)$ where $V^\mathbf G$ is a finite set and $E^\mathbf G \subseteq V^\mathbf G\times V^\mathbf G$.
  A \emph{homomorphism} of such digraphs $f\colon \mathbf G \to \mathbf H$ is a mapping $f\colon V^\mathbf G \to V^\mathbf H$ that preserves edges in the sense that if $(u, v) \in E^\mathbf G$ then $(f(u), f(v)) \in E^\mathbf H$.

  Such digraphs correspond to copresheaves over the category $\cat D$ with two objects and two non-trivial arrows:
  \[\begin{tikzcd}
    E \arrow[r, shift left, "s"] \arrow[r, shift right, "t"'] & V
  \end{tikzcd}\]
  For a given finite digraph $\mathbf G = (V^\mathbf G, E^\mathbf G)$, define a copresheaf $G\colon \cat D \to \fin$ with $G(V) = V^\mathbf G$ and $G(E) = E^\mathbf G$, where the map $G(s)\colon E^\mathbf G \to V^\mathbf G$ is the first projection and $G(t) \colon E^\mathbf G \to V^\mathbf G$ is the second projection. We denote the resulting copresheaf by $\mathbf G^\mkPrsh$. Clearly, the assignment $\mathbf G \mapsto \mathbf G^\mkPrsh$ is a functor.

  Copresheaves over $\cat D$ are (slightly) more general than (simple) digraphs in the sense that parallel edges are allowed, and, moreover, homomorphisms (natural transformation) need to specify the image of each edge. In combinatorics, such structures are usually called \emph{multidigraphs}.

 Nevertheless, given any copresheaf $G\colon \cat D \to \fin$, we may define a digraph $\mathbf G = G^\mkStr$ with $V^\mathbf G = G(V)$ and
  \[
    E^\mathbf G = \{ (u, v) \mid \exists e \in G(E),
      G(s)(e) = u \text{ and } G(t)(e) = v \}.
  \]
  Note that $\mathbf G^{\mkPrsh\mkStr}$ is isomorphic to $\mathbf G$, but $G^{\mkStr\mkPrsh}$ is not isomorphic to $G$---nevertheless, they are still \emph{homomorphically equivalent}, i.e.\ there is a pair of homomorphisms $G \rightleftarrows G^{\mkStr\mkPrsh}$. Since CSPs are defined in terms of existence of a homomorphism, this means that from the CSP perspective, the two structures are indistinguishable.
\end{example}

\begin{example}[Graphs as copresheaves]
  A \emph{graph} is traditionally defined as a pair $\mathbf G = (V^\mathbf G, E^\mathbf G)$ where $E^\mathbf G \subseteq \binom V2$, i.e.\ $E^\mathbf G$ is a set of two-element subsets of $\mathbf V$. In the context of CSPs, they are often treated as digraphs where the relation $E^\mathbf G \subseteq V^\mathbf G \times V^\mathbf G$ is symmetric (note that this implicitly allows self-loops, i.e.\ edges from $u$ to $u$).

  Graphs (with self-loops) naturally correspond to copresheaves on the category $\cat U$ with three non-trivial morphisms
  \[\begin{tikzcd}
    E \arrow[r, shift left, "s"]
      \arrow[r, shift right, "t"']
      \arrow[loop left, "r"]
    & V
  \end{tikzcd}\]
  where $r^2 = \id_E$, $sr = t$, and $tr = s$. The operation $r$ is interpreted as the flipping of the orientation of an edge.
  Similarly as for digraphs, the copresheaves over $\cat U$ correspond to \emph{multigraphs} where parallel edges between given two vertices are allowed.

  The benefit of our approach is that we can now treat the category of multigraphs (copresheaves on $\cat U$) as a first-class citizen of our theory and not as a subcategory of digraphs.
\end{example}

\begin{example}[Relational structures as copresheaves]
    \label{ex:structs-as-presheaves}
    Given a finitary relational signature \(\Sigma\), the \emph{signature category} \(\cat R_\Sigma\) has one object for each relational symbol \(R\in \Sigma\) and an additional object \(V\) representing the set of vertices. Furthermore, the only non-trivial morphisms are the projections \(\hom(R, V) = \{ p_1^R, \dots, p_{r}^R\}\) where \(r\) is the arity of \(R\).

    As in Example~\ref{ex:graphs-as-presheaves} we have that every relational structure \(\mathbf A = (V^\mathbf A, (R^\mathbf A)_{R\in\Sigma})\) induces the copresheaf \(\mathbf A^\mkPrsh \colon \cat R_\Sigma \to \fin\) and any copresheaf \(A\colon \cat R_\Sigma \to \fin\) induces the corresponding relational structure \(A^\mkStr\) without duplicit tuples in any of the relations.
    As before, we still maintain that \(\mathbf A \cong \mathbf A^{\mkPrsh\mkStr}\) and~\(A\rightleftarrows A^{\mkStr\mkPrsh}\).
\end{example}

\subsection{Yoneda embedding}

For a finite category $\cat S$, the \emph{Yoneda embedding} is a functor
\[
  \yo\colon \cat S^\op \to \pshcat S
\]
defined as $s \mapsto \hom(s, {\ARG})$.\footnote{The symbol $\yo$ is the hiragana character for `yo'.} As a consequence of the \emph{Yoneda lemma}, this functor is a full embedding. More precisely, the Yoneda lemma states that for every \(X\colon \cat S \to \fin\) and \(s\in \cat S\),
\[
  \hom(\yo(s), X) \simeq X(s)
\]
and that this isomorphism is natural in $s$ and $X$ where, in this case, \(\hom(Y, X)\) is the collection of natural transformations~\(Y \to X\).

\begin{example}[Example~\ref{ex:graphs-as-presheaves} cont.]
  \label{ex:yo-structures}
  Let us describe the Yoneda embedding $\cat D^\op \to \pshcat D$.
  We start by describing the values of $\yo$ on the two objects of $\cat D$.
  By definition, $\yo(d) = \hom_{\cat D}(d, {-})$ are functors $\cat D\to\fin$, i.e.\ directed multigraphs.
  
  Putting $d = V$, the graph $\yo(V)$ has precisely one vertex, since $\hom(V,V) = \{\id_V\}$, and no edges, since $\hom(V,E)$ is empty.
  The graph $\yo(E)$ has vertices $\hom(V, E) = \{s, t\}$, and has an edge corresponding to $\id_E$ that connects $s = s\circ \id_E$ and $t = t\circ \id_E$. In plain words, the graph $\yo(E)$ is an oriented edge from $s$ to $t$.
  
   Lastly, $\yo(s) \colon \yo(V) \to \yo (E)$ corresponds to the homomorphism that maps the unique vertex of $\yo(V)$ to $s$, and $\yo(t)\colon \yo(V) \to \yo(E)$ corresponds to the homomorphism that maps the unique vertex to~$t$.
\end{example}

\subsection{Limits and the Label Cover problem}

In combinatorics, the uniform CSP, which we introduced in Definition~\ref{def:uniform-csp}, is often called \emph{Label Cover}.
As we mentioned in the introduction, the uniform CSP is NP-complete; see also the note below Corollary~\ref{cor:bulatov-zhuk-NP-complete}.
We rephrase the definition in more categorical language for easier reference.

\begin{definition}[Uniform CSP]
\label{def:CSP}
    Given a finite category $\cat J$ and a functor $D\colon \cat J\to \fin$, decide between the cases:
    \begin{description}
    \item[\Yes] if $D$ admits a solution, i.e.\ $\lim D\neq \emptyset$; and
    \item[\No] if $D$ does not admit a solution, i.e.\ $\lim D=\emptyset$
    \end{description}
\end{definition}

The inputs for this problem come from the category \(\fcat \comma \fin\) where \(\fcat\) is the (1-\nobreak)\allowbreak category of finite categories and functors between them, and \(\fcat \comma \cat S\) is the comma category whose objects are functors \(D\colon \cat J \to \cat S\) and morphisms \(D \to D'\) are functors \(H\colon \cat J \to \cat J'\) such that \(D = D' \circ H\).

A \emph{solution} to $D\colon \cat J \to \fin$ is an assignment $d\in \prod_{i\in \cat J}$, such that, for all $f\colon i\to j$, we have $D(f)(d_e) = d_j$.
Observe that the set of solutions (together with the projections $d\mapsto d_e$) is the categorical limit of $D$ in $\fin$. We will henceforth identify the set of solutions to $D$ with $\lim D$.

We conclude this section with a brief note on the computational complexity of colimits in $\fin$. Its computational difficulty comes from computing the symmetric transitive closure of a binary relation, which can be done in log-space due to a result of Reingold~\cite{Reingold08}.

\begin{lemma} \label{lem:reingold}
    Colimits in $\fin$ are computable in log-space. More precisely, there is a log-space algorithm which on input gets a finite diagram $D\colon \cat J \to \fin$, and outputs a set $L$ such that $L = \colim_{i\in \cat J} D(i)$.
\end{lemma}

\section{Two versions of promise-CSPs}
    \label{s:pcsp-basics} \label{sec:two-pcsps}

In this subsection, we give two formal definitions of \emph{promise constraint satisfaction problems (PCSPs)}.
A \emph{promise (decision) problem} over some class $C$, whose elements are called \emph{instances}, is a pair of subclasses $Y, N \subseteq C$; where the elements of $Y$ are \emph{\Yes{} instances} and the elements of $N$ are \emph{\No{} instances}. We usually assume that $Y$ and $N$ are disjoint. While $Y$ and $N$ might not be complementary, the ``promise'' is that any given $c\in C$ is either in $Y$ or in $N$. A \emph{reduction} from one such problem $Y, N\subseteq C$ to another problem $Y', N'\subseteq C'$ is a function $f\colon C \to C'$ that preserves \Yes{}  and \No{} instances, i.e.\ such that $f(Y) \subseteq Y'$ and $f(N) \subseteq N'$, which we call \emph{soundness} and \emph{completeness}, respectively. Of particular interest to us are reductions $f$ which can be computed in log-space, in which case we say that $f$ is a log-space reduction.

\subsection{Homomorphism perspective}
We define PCSPs as a promise version of the homomorphism problem (Definition~\ref{def:csp-v1}) in an arbitrary category, although usually we consider categories of finite copresheaves.

\begin{definition}[Promise constraint satisfaction problems]
    \label{def:pcsp-v1}
  Fix a category $\cat C$. A \emph{promise template} is a pair of objects $(A, B)$ of $\cat C$ such that there exists a morphism $A \to B$, which ensures that the \Yes{} and \No{} cases are disjoint. The \emph{promise constraint satisfaction problem} $\hPCSP(A, B)$ is the promise problem defined as follows.
  Given an object $X$ of $\cat C$, output:
  \begin{description}
    \item[\Yes] if there is a morphism \(X \to A\); and
    \item[\No] if there is no morphism $X \to B$.
  \end{description}
\end{definition}

If $A = B$ then the \Yes{} and \No{} instances of $\hPCSP(A, B)$ are complementary. In that case, we denote the problem by $\hCSP(A)$, i.e.\ $\hCSP(A) = \hPCSP(A, A)$.

We use the term \emph{finite-template PCSP} for PCSPs in a category of finite copresheaves, i.e.\ when $\cat C = \pshcat S$ for a finite category~$\cat S$.
The finiteness ensures that the inputs can be encoded in a data structure taking a finite amount of memory, and hence we can talk about computational aspects of the problems. Occasionally, it is useful to talk about PCSPs over other categories, such as $[\fin,\fin]$, in which case we do not mention the computational aspects.
It may be observed that finite-template PCSPs are always in \NP. A~focus of the study of these PCSPs is to identify problems that are \emph{tractable}, i.e.\ to find an algorithm that decides in polynomial time in the size of an input instance if it is a \Yes{} or \No{} instance (and it is guaranteed/promised that it is one of the two).

\begin{example}[Example~\ref{ex:graphs-as-presheaves} cont.]
\label{ex:3-col}
Let $\cat S \coloneq \cat D$ be the category from Example~\ref{ex:graphs-as-presheaves}. If $A,B\in\pshcat S$ are two copresheaves, then the problem $\hPCSP(A,B)$ is a homomorphism problem of multidigraphs: Given a multidigraph $X\in\pshcat S$, decide whether there exists a homomorphism $X\to A$ or not even a homomorphism $X\to B$.

In particular, if $K_3$ is the copresheaf corresponding to the $3$-clique, i.e.\ with $K_3(V) = \{0, 1, 2\}$ and $K_3(E) = \{ 01, 10, 12, 21, 02, 20 \}$ with $K_3(s)$ and $K_3(t)$ being the projections on the first and second coordinates respectively, the problem $\hCSP(K_3)$ is equivalent to graph $3$-colouring.
\end{example}

\subsection{Satisfiability perspective}

A curious category theorist might ask what the problem is that corresponds to the PCSP problem from Example~\ref{ex:3-col} via the Grothendieck construction.
In case \(A = B\), this turns out to correspond to a restriction of the Universal CSP (Definition~\ref{def:CSP}).
The resulting problem $\sCSP(A)$ is parameterised by a finite copresheaf $A\colon \cat S\to\fin$
and mandates that input constraints must already appear in $A$, i.e.\ constraints must be maps of the form $A(g)\colon A(s)\to A(s')$ for some morphism $g\colon s\to s'$ in $\cat S$.
The promise version of this problem is defined as follows.

\begin{definition}[PCSP, version 2]
\label{def:pcsp-v2}
    Let \((A,B)\) be a promise template where $A, B\colon \cat S \to \fin$ are copresheaves.
  We define $\sPCSP(A, B)$ as the following promise problem.
  Given a finite category $\cat J$ and a functor $D \colon \cat J \to \cat S$ decide between the cases:
  \begin{description} \setlength{\itemindent}{-1em}
    \item[\Yes] if $A \circ D$ admits a solution; and
    \item[\No] if $B \circ D$ does not admit a solution.
  \end{description}
    Observe that the \Yes{} and \No{} cases are disjoint, as the required natural transformation \(h\colon A \to B\) induces a map between the limits.
    \[
      \lim(A\circ D)     \to     \lim(B\circ D), \qquad
      (a_i)_{i\in\cat J} \mapsto \big(h_{D(i)} (a_i) \big)_{i\in \cat J}
    \]
\end{definition}

\begin{example}
    Fix $d \geq 1$, and let $\cat S$ be the subcategory of $\fin$ consisting of all objects, but only those maps which are at most $d$-to-1.
    If $A\colon \cat S\to \fin$ is the inclusion functor, then $\sCSP(A)$ is the problem called \emph{$d$-to-1 Games}; see, e.g.\ \cite[Definition 1.2]{Dinuretal25}.
    There are several conjectures about the hardness of approximability of these problems; the inapproximability of 1-to-1 Games is the famous \emph{Unique Games Conjecture} of Khot \cite{Khot02}, and Dinur et al.~\cite{Dinuretal25} provides some recent results on a closely related \emph{2-to-1 Conjecture}.
\end{example}

\begin{remark} \label{ex:pp}
  By definition, $\lim (A\circ D)$ is non-empty whenever ``\emph{there exists} a tuple \((a_i)_{i\in \cat J}\) such that, \emph{for every} morphism \(f\colon i \to j\) in \(\cat J\), \(A(D(f))(a_i) = a_{j}\)''.
  This statement can be rewritten as satisfaction of a first-order sentence in $A$, where we view $A$ as a multisorted structure (with one sort for each object of $\cat S$) with a unary function symbol for each morphism in $\cat S$. Moreover, by replacing the \emph{for all} above with a big conjunction, we obtain a sentence
  \begin{equation}
    \exists (a_i \in A(D(i)))_{i\in \cat J}\,
    \bigwedge_{i, j\in \cat J}\bigwedge_{f\colon i \to j} A(D(f))(a_i) = a_j
    \label{eq:pp-from-functor}
  \end{equation}
  which uses only existential quantifiers, conjunctions, and atoms. Such sentences are called \emph{primitive positive}.
\end{remark}

\begin{example}
  Let $\cat S \coloneq\cat D$ be the category form Example~\ref{ex:graphs-as-presheaves}. Then each $f\colon i\to j$ is sent to either $s, t$ or an identity morphism. The atom $A(D(f))(a_i) = a_j$, in case $D(f) = s$, expresses that the source of the edge $a_i\in A(E)$ is the vertex $a_j\in A(V)$, and similarly, in case $D(f) = t$, the atom expresses that the target of the edge $a_i$ is the vertex $a_j$.
  In particular, the big conjunction asks for the existence of a tuple $(a_i)_{i \in \cat J}$ of edges and vertices in $A$ which are related as required.
  This can be viewed as the satisfaction of a primitive positive sentence in a multidigraph $A$ interpreted as a two-sorted structure with a sort for vertices and a sort for edges.
\end{example}

\begin{remark}[Remark~\ref{ex:pp} cont.]
  The observation in the previous remark provides a reduction from $\sCSP(A)$ to the problem of satisfaction of a primitive positive sentences in $A$; cf.~Appendix~\ref{sec:translations}.
  Conversely, every primitive positive sentence~$\varphi$ with atoms of the form $A(f')(x) = y$ yields a functor $D \colon \cat J \to \cat S$ such that $A$ satisfies $\varphi$ if and only if $\lim(A\circ D)$ is non-empty.\footnote{Note that equality $x = y$ can be expressed as $A(\id)(x) = y$, and atoms of the form $A(f)(A(g)(x)) = y$ can be expressed as $A(f\circ g)(x) = y$.}
  The objects of $\cat J$ are the variables of $\varphi$, and $D$ maps each variable to the corresponding sort, which is an object of $\cat S$.
  Each atom of the form $A(f')(x) = y$ induces a morphism $x\to y$ in $\cat J$ mapped by $D$ to~$f'$.
  However, the resulting $\cat J$ need not be a category since its morphisms might not compose.
  This can be dealt with by, e.g., adding a copy $x'$ of each variable $x$ in $\varphi$ and adding a morphism $x' \to x$ in $\cat J$ which maps to the identity. Then we use $x'$ for any outgoing arrows from $x$ and keep $x$ only for incoming arrows, to ensure that $\cat J$ has no non-trivial composable arrows.

  Hence, the problem $\sCSP(A)$ is equivalent to the satisfaction problem of primitive positive sentences in $A$.
\end{remark}

\subsection{Equivalence via Grothendieck}
We now relate the two versions of PCSPs via the \emph{discrete Grothendieck construction}. Not only does this provide a powerful theoretical tool for later sections, it also proves that the two PCSPs are equivalent up to log-space reductions.

\begin{theorem}
\label{thm:hom=fib}
    Let $(A, B)$ be a promise template where \(A, B\colon \cat S \to \fin\) are finite copresheaves. Then $\hPCSP(A, B)$ and $\sPCSP(A, B)$ are log-space equivalent.
    More precisely, these reductions are given by the discrete Grothendieck construction and its left adjoint, as shown below.
    \[
    \begin{tikzcd}
    {\hPCSP(A,B)} \arrow[d, "\dashv", phantom] \arrow[d, "\gr", shift left=3] \\
    {\sPCSP(A,B)} \arrow[u, "\gl{\ARG}", shift left=3]
    \end{tikzcd}
    \]
\end{theorem}

Next, we introduce the aforementioned constructions and prove a few basic properties about them.

\begin{definition}[Grothendieck construction]
    Let $X\colon \cat S \to \fin$ be a copresheaf. The \emph{Grothendieck construction} \(\gr X\) is a functor of type
    \[
    \gr X\colon \textstyle\el X \to \cat S
    \]
    where $\el X$ is the category of \emph{elements} of $X$, consisting of pairs $(s, x)$ such that $s \in \cat S$ and $x\in X(s)$. The morphisms $(s, x) \to (t, y)$ in \(\el X\) are given by morphisms $f\colon s \to t$ in $\cat S$ such that $X(f)(x) = y$.    The functor $\gr X\colon \el X \to \cat S$ is then the first projection functor, \((s,x)\mapsto s\). Notice that if \(\cat S\) is finite, then so is \(\el X\).
\end{definition}

\begin{example} \label{ex:elements}
    Consider the graph $G$ consisting of three vertices $0,1,2$ and two edges $3,4$ arranged in sequence, depicted in the figure below on the left.
    \[
        \begin{tikzpicture}[fatnode/.style = {circle, fill, inner sep = 0,
                              outer sep = 1pt, minimum size = 4pt},
                            baseline = (base.base) ]
          \node [fatnode] (0) [label=below:$0$] {};
          \node (base) [anchor=north] at (0.south) {$\phantom 0$};
          \node [fatnode] (1) [label=below:$1$] at (1, 0) {};
          \node [fatnode] (2) [label=below:$2$] at (2, 0) {};
          \draw[->] (0) -- (1) node[midway, above] {3};
          \draw[->] (1) -- (2) node[midway, above] {4};
        \end{tikzpicture}
        \qquad\qquad
        \begin{tikzcd}[column sep=1em, row sep=1em,
                       baseline = (\tikzcdmatrixname-2-1.base)]
          & (E,3) \arrow[ld, "s"'] \arrow[rd, "t"] &
          & (E,4) \arrow[ld, "s"'] \arrow[rd, "t"] \\
          (V,0) & & (V,1) & & (V,2)
        \end{tikzcd}
    \]
    Under the correspondence from Example~\ref{ex:graphs-as-presheaves}, we view $G$ as a copresheaf $\cat D\to \fin$ with $G(V)=\{0,1,2\}$, $G(E)=\{3,4\}$ and the appropriate maps $G(s), G(t)$. The category of elements $\el G$ then consists of five objects and four morphisms as depicted in the figure above on the right.
\end{example}

\begin{lemma}[$\gr$ is a reduction]
\label{lem:gr1}
  For copresheaves $X, A\colon \cat S \to \fin$, there is a bijection
  \[
    \hom(X, A) \cong \lim(A\circ \gr X).
  \]
  In particular, the functor $\gr\colon \pshcat S\to\fcat \comma \cat S$ is a reduction from $\hPCSP(A, B)$ to $\sPCSP(A, B)$ for any template \((A, B)\).
\end{lemma}

\begin{proof}
    The bijection can be defined explicitly.
    Given a natural transformation $h\colon X\to A$, define an element $a\in\lim (A\circ \gr X)$ as $a_{(s,x)}\coloneq h_s(x)$ for $s\in\cat S$ and $x\in X(s)$. To verify that this is indeed a solution of $A\circ \gr X$, take $g\colon s\to t$ in $\cat S$ with $X(g)(x) = y$, and compute
    \[
        A(g)(\alpha_{(s,x)}) = A(g)(h_s(x)) = h_t(X(g)(x)) = h_t(y) = \alpha_{(t,y)}
    \]
    where the second equality is due to the naturality of $h$. Similarly, one can show that if $a$ is an element of the limit, then $h$ defined as $h_s(x)\coloneq a_{(s,x)}$ is a natural transformation $X\to A$.
\end{proof}

The Grothendieck construction admits a left adjoint; see, e.g., \cite[Section 4.7]{Kelly82}, which provides the second reduction.
Recall that a category is \emph{locally finite} if its hom sets are finite.

\begin{definition}[Grothendieck's left adjoint]
    Fix a locally finite category~$\cat S$ and consider a functor $D\colon\cat J\to \cat S$ for a finite category~$\cat J$. Define a copresheaf $\gl D\in\pshcat S$ as follows.
    \[
    \gl D = \colim_{\cat J^\op}  ( \yo \circ D^\op )
    \]
    where $\yo$ be the Yoneda embedding $\yo\colon \cat S^\op\to [\cat S,\fin]$.
\end{definition}

\begin{figure*}
\begin{tikzpicture}
      [fatnode/.style = {circle, fill, inner sep = 0, outer sep = 1pt, minimum size = 4pt},decoration=snake,
   line around/.style={decoration={pre length=#1,post length=#1}}, scale= 0.8]
      \node (-6) at (-6, 0) {E};
      \node (-5) at (-5, 1) {E};
      \node (-4) at (-4, 0) {V};
      \node (-3) at (-3, 1) {E};
      \node (-2) at (-2.5, 0.5) {};
      \node (-1) at (-0.5, 0.5) {};
      \draw[->] (-5) -- (-6) node[midway, above left] {$\id$};
      \draw[->] (-5) -- (-4) node[midway, below left] {$t$};
      \draw[->] (-3) -- (-4) node[midway, below right] {$s$};
      \draw[->, decorate] (-2) -- (-1) node[midway, above=3pt] {$\yo$};
      \node [fatnode] (0) {};
      \node [fatnode] (1) at (1, 0) {};
      \node [fatnode] (2) at (3, 0) {};
      \node [fatnode] (3) at (1, 1) {};
      \node [fatnode] (4) at (2, 1) {};
      \node [fatnode] (5) at (4, 1) {};
      \node [fatnode] (6) at (5, 1) {};
      \node at  (1, 0.5) {$\id$};
      \draw[->] (0) -- (1);
      \draw[->, dashed, shorten > = 5pt, shorten < = 5pt]
      (0) -- (3);
      \draw[->, dashed, shorten > = 5pt, shorten < = 5pt]
      (1) -- (4);
      \draw[->, dashed, shorten > = 5pt, shorten < = 5pt]
      (2) -- (4) node[midway, xshift=-6pt, yshift=-6pt] {$\yo(t)$};
      \draw[->, dashed, shorten > = 5pt, shorten < = 5pt]
      (2) -- (5) node[midway, xshift=6pt, yshift=-6pt] {$\yo(s)$};
      \draw[->] (3) -- (4);
      \draw[->] (5) -- (6);
      \node (7) at (5.5, 0.5) {};
      \node (8) at (7.5, 0.5) {};
      \node [fatnode] (9) at (8, 0.5) {};
      \node [fatnode] (10) at (9, 0.5) {};
      \node [fatnode] (11) at (10, 0.5) {};
      \draw[->] (9) -- (10);
      \draw[->] (10) -- (11);
      \draw[->, decorate] (7) -- (8) node[midway, above=3pt] {$\colim$};
\end{tikzpicture}
\caption{Illustration of the left adjoint to the Grothendieck construction.}
\label{fig:gl}
\end{figure*}

\begin{example}[see Figure~\ref{fig:gl}]
\label{ex:gl}
Let $\cat D$ be the category of Example~\ref{ex:graphs-as-presheaves}, and $\cat J$ be the category with four objects and three morphisms arranged in a zig-zag pattern. Let $D\colon \cat J\to \cat D$ be the diagram depicted on the left-hand side of Figure~\ref{fig:gl}.
Then $\gl{D}$ is computed by first applying the Yoneda embedding, to obtain a diagram $\cat J^\op\to\pshcat D$, and then taking its colimit.

\end{example}

\begin{lemma}[$\gl\ARG$ is a reduction]
    \label{l:gl-reduction}
    Fix a copresheaf \(A\colon \cat S \to \fin\) where \(\cat S\) is locally finite.
    Then for every finite category $\cat J$ and any functor $D\colon\cat J\to\cat S$ there is a bijection
    \[
      \lim (A\circ D)\cong \hom(\gl D,A).
    \]
    In particular, the functor $\gl\ARG\colon \Cat\comma \cat S\to \pshcat S$ is a reduction from $\sPCSP(A,B)$ to $\hPCSP(A,B)$ for any template \((A,B)\).
\end{lemma}

\begin{proof}
    Let $\yo$ be the Yoneda embedding $\cat S^\op\to\pshcat S$ and compute the following.
    \begin{align*}
    \lim_{i\in\cat J} A( D(i)) &\cong
    \lim_{i\in \cat J} \hom ( \yo( D(i)), A )
        \cong \hom ( \colim_{i\in\cat J^\op} ( \yo (D(i) ), A ) = \hom(\gl D, A)
    \end{align*}
The first bijection follows from the Yoneda lemma, while the second is due to the fact that $\hom$-functors preserve limits.
\end{proof}

To finish the proof of Theorem~\ref{thm:hom=fib}, we need to show that in the case where $\cat S$ is finite, both $\gr$ and $\gl\ARG$ are computable in log-space, which is clear for $\gr$. For $\gl\ARG$ observe the only computational challenge is computing colimits in $\pshcat S$, which can be done in log-space due to Lemma~\ref{lem:reingold}.

\begin{remark}[Remark~\ref{ex:pp} cont.]
    Lemmas~\ref{lem:gr1} and~\ref{l:gl-reduction} can be viewed as a categorical formulation of the Chandra--Merlin correspondence~\cite{chandra1977optimal} for primitive positive sentences. We remark that Bonchi, Seeber, and Sobociński~\cite{BonchiSeeberSobocinski18} provide a similar categorical treatment of the correspondence using string diagrams.
\end{remark}

\section{Kan extensions provide reductions}
  \label{sec:square}

One of the key achievements of the algebraic approach to CSPs is a sufficient criterion for the existence of a log-space reduction between two promise CSPs. It was first provided in the scope of finite-template CSPs by Bulatov, Jeavons, and Krokhin \cite{BulatovJK05}, and in the scope of PCSPs by Barto, Bulín, Krokhin, and Opršal \cite[Theorem 3.1]{BBKO21}. Both of these results are formulated using the notion of \emph{polymorphisms}. We phrase this criterion categorically in terms of right Kan extensions (see Definition~\ref{def:ran}). This statement is equivalent with the usual polymorphism phrasing by Lemma~\ref{lem:pol-is-ran} below.

\begin{theorem}
  \label{thm:ranGivesReductions}
    Consider promise templates \((A,B)\) and \((A',B')\) consisting of finite copresheaves $A, B \colon \cat S \to \fin$ and $A', B'\colon \cat T \to \fin$.
    If there is a natural transformation $\Ran_{A'} B'\to\Ran_A B$, then there is a log-space reduction from $\hPCSP(A,B)$ to $ \hPCSP(A',B')$.
\end{theorem}

\subsection{Kan extensions and polymorphism minions}
We recall the definition of Kan extensions in the special case we require, namely, for (finite) copresheaves.

\begin{definition}\label{def:ran}
    Given a copresheaf $A\colon\cat S \to\fin$, one can define the precomposition functor
\[
    [\fin,\fin] \to     \pshcat S, \qquad
    M           \mapsto M\circ A.
\]

The \emph{right Kan extension along} $A$ is a functor $\Ran_A\colon\pshcat S\to[\fin,\fin]$ which is right adjoint to the precomposition functor. Explicitly, for any two functors $M\in[\fin,\fin]$ and $B\in\pshcat S$, there is a bijection natural in both $M$ and $B$
\begin{equation}
    \hom(M\circ A, B)\cong \hom(M,\Ran_A B).
    \label{eq:ran}
\end{equation}

\noindent
Dually, the \emph{left Kan extension} is defined by a natural bijection
\begin{equation}
    \hom(\Lan_A B,M) \cong \hom(B, M\circ A).
\end{equation}
\end{definition}

These universal properties define Kan extensions uniquely up to isomorphism.
In the following, we give an explicit description of the Kan extension in Theorem~\ref{thm:ranGivesReductions}.

\begin{lemma} \label{lem:pol-is-ran}
    Given two copresheaves $A, B\colon \cat S\to \fin$ where \(\cat S\) is finite,
    the right Kan extension $\Ran_A B$ exists. Furthermore, there is a bijection
    \[
    \Ran_A B( N )  \cong \hom(A^N,B)
    \]
    for any finite set $N$, where $A^N$ is the component-wise $N$-th power of~$A$. This bijection is natural in both $N$ and $B$.
\end{lemma}

\begin{proof}
   Recall that the Yoneda-embedding $\yo\colon \fin^\op\to [\fin,\fin]$ is the mapping $N\mapsto (-)^N$. Therefore, we compute the following
   \[
       \Ran_A B (N) \cong
       \hom( \yo(N), \Ran_A B) \cong
       \hom( \yo (N) \circ A, B) =
       \hom( A^N, B )
   \]
   The first bijection is the Yoneda lemma, and the second is \eqref{eq:ran}.
\end{proof}

The above lemma links right Kan extensions with several core notions of the algebraic approach, namely with the notions of \emph{polymorphisms} and \emph{polymorphism minions}.
Given objects \(A,B\) in some category, the functor $\pol(A, B)\colon \fin \to \set$ defined by $N \mapsto \hom(A^N, B)$ is called the \emph{polymorphism minion}; in the case $A = B$, a simpler notation $\pol(A)$ is used instead of $\pol(A, A)$.
An \emph{(abstract) minion} is a term used by algebraists for a functor $\fin \to \set$ and \emph{polymorphisms} are homomorphisms $A^n \to B$ for some \(n\).
Since in this paper, we work with locally finite categories, we may retype $\pol(A, B)$ as a functor $\fin\to \fin$. Then Lemma~\ref{lem:pol-is-ran} shows that, for two finite copresheaves $A$ and $B$, \(\pol(A, B) \cong \Ran_A B\).

We note that the above proof shows that $\pol(A, \ARG)$ is a right adjoint to $\ARG\circ A$. The left adjoint to $\pol(\mathbf A, \ARG)$, where $\mathbf A$ is a relational structure, was discussed in \cite[Lemma 4.4]{BBKO21}; it is often called the \emph{free structure} and denoted by \(\freeStr {\mathbf A} M\) where $M\colon \fin \to \fin$.  It is a conceptualisation of a well-known universal algebraic trick dating back to Maltsev~\cite{Maltsev54}; the definition first appeared in \cite[Section 7]{BartoOP18} in the context of CSPs and in \cite[Definition 4.1]{BBKO21} in the context of PCSPs. The latter definition is given in the language of `function minions' and is a very concrete algebraic construction; see also \cite[Definition A.2]{BrakensiekGWZ20} for a construction of free structures from abstract minions. We explain the construction in the following example.

\begin{example}
    \label{ex:free-structure}
    Let $\mathbf A$ be a relational structure (cf.\ Example~\ref{ex:structs-as-presheaves}), and let $M = \pol(A', B')$ be a polymorphism minion. Following \cite[Definition 4.1]{BBKO21} we assume that the universe of $\mathbf A$ is the set $[n] = \{1, \dots, n\}$.
    The free structure $\freeStr{\mathbf A} M$ is defined as follows:
    The domain is \(\hom(A'^n, B')\) and, for an \(k\)-ary relation symbol \(R\in \Sigma\), we view projections \(A(p^R_i)\colon A(R) \to A(V)\) as mappings \(\rho_i\colon [m] \to [n]\) where \([m]\) is some enumeration of \(R^\mathbf A\).
    Then, the interpretation of~\(R\) in~\(\freeStr {\mathbf A} M\) consists of tuples \((r_1, \dots, r_k)\in \freeStr {\mathbf A} M^k\) for which there exists \(r\in M([m])\), i.e.\ a homomorphism \(r\colon B^{[m]} \to B'\), such that \(r_i(b_1,\dots,b_n) = r(b_{\rho_i(1)}, \dots, b_{\rho_i(m)})\) for each $i$ and all $b_1, \dots, b_m \in B$. In other words, we have \(r_i = A(p^R_i)(r)\) for all~$i$.
    Consequently, we observe that \(\freeStr{\mathbf{A}} M\) is the relational structure obtained as \((M\circ \mathbf A^\mkPrsh)^\mkStr\).
\end{example}

\subsection{Equivalence via Kan}
As a preparation for Theorem~\ref{thm:ranGivesReductions} we prove the equivalence of the computational homomorphism problem with a non-com\-pu\-ta\-tion\-al satisfaction problem, which concerns functors \(\fin \to \fin\).

\begin{theorem}
    \label{t:fundamental-theorem}
Consider a promise template $(A,B)$ where $A,B$ are finite copresheaves. Then there is a reduction between the following two problems.
\[
\begin{tikzcd}
{\hPCSP(A,B)} \arrow[r, "\rho_A", shift left] & {\sPCSP(\id,\Ran_A B)} \arrow[l, "\pi_A", shift left]
\end{tikzcd}
\]
Moreover, $\rho_A$ is computable in log-space. On the other hand, $\pi_A$ is also computable in log-space, if one restricts it to inputs $D\colon\cat J\to \fin$ where the sets $D(i)$ are of bounded size.
\end{theorem}

For the proof, one examines the following diagram and notices that all arrows are reductions between the corresponding PCSPs:
\begin{equation}
\begin{tikzcd}
\hPCSP(A,B)
    \arrow[d, "\gr", shift left=2]
    \arrow[d, "\dashv", phantom]
&
\hPCSP(\id,\Ran_A B)
    \arrow[d, "\gr", shift left=2]
    \arrow[d, "\dashv", phantom]
    \arrow[l, "\ARG\circ A"']
\\
\sPCSP(A,B)
    \arrow[u, "\gl\ARG", shift left=2]   \arrow[r, "A\circ \ARG"']
&
\sPCSP(\id,\Ran_A B)
    \arrow[u, "\gl\ARG", shift left=2]
\end{tikzcd}
\tag{$\square$} \label{eq:square}
\end{equation}
We showed that $\gr$ and $\gl\ARG$ are reductions in Theorem~\ref{thm:hom=fib} and, since~$\cat S$ is finite, $\gr$ on the left hand side can be computed in log-space.

\begin{lemma} \label{lem:4.6}
    The functor $D\mapsto A\circ D$ is a valid reduction from $\sPCSP(A,B)$ to  $\sPCSP(\id, \Ran_A B)$, and is computable in log-space.
\end{lemma}

\begin{proof}
    To check that the reduction is valid, let $D\colon \cat J\to\cat S$ be an input to $\sPCSP(A,B)$.
    \begin{description}
        \item[Soundness:]
        If $A\circ D$ has a solution, then so does $\id \circ A\circ D$.
        \item[Completeness:]
        The counit $\Ran_A B\circ A\to B$ of the adjunction~\eqref{eq:ran} induces a map
        \[
        \lim (\Ran_A B\circ A\circ D )\to \lim (B\circ D).
        \]
        Thus, if the former set is non-empty then so is the latter. \qedhere
    \end{description}
\end{proof}

\begin{lemma}
    \label{l:precomp-reduction}
    The functor $M\mapsto M\circ A$ is a valid reduction from $\hPCSP(\id,\Ran_A B)$ to $\hPCSP(A,B)$.
\end{lemma}

\begin{proof}
    Similarly as before, let $M\colon \fin\to\fin$ be an input to $\hPCSP(\id,\Ran_A B)$.
    \begin{description}
    \item[Soundness:] If $M\to \id$, then also $M\circ A\to \id\circ A = A$.
    \item[Completeness:] If $M\to \Ran_A B$, then also $M\circ A\to B$ by \eqref{eq:ran}. \qedhere
    \end{description}
\end{proof}

We a priori cannot talk about computational aspects of the right side of \eqref{eq:square} since functors $M\colon \fin \to\fin$ are not finitely encodable. However, the composition $(\ARG\circ A)\circ\gl\ARG$ sends functors $D\colon\cat J\to\fin$ for some arbitrary finite $\cat J$ to functors $\cat S\to \fin$ in the following way. Since the Yoneda embedding $\yo\colon\fin^\op\to[\fin,\fin]$ can be expressed as $N\mapsto (\ARG)^N$,
\[
  D\mapsto \colim(\yo\circ D^\op)\circ A = \colim_{i\in\cat J} (\ARG)^{D(i)}\circ A = \colim_{i\in\cat J} A^{D(i)}
\]
This computation is clearly exponential in the size of the sets $D(i)$. However, if the size of the image of $D$ is bounded from above by some constant $k$, then the only computational challenge  is to compute the colimit,  which can be done in log-space due to Lemma~\ref{lem:reingold}. This completes the proof of Theorem~\ref{t:fundamental-theorem} and, consequently, also of Theorem~\ref{thm:hom=fib}.
Finally, we turn to the proof of the main theorem of this section.

\begin{proof}[Proof of Theorem~\ref{thm:ranGivesReductions}.]
The reduction from $\hPCSP(A,B)$ to $ \hPCSP(A',B')$ is summarised in the following diagram.
\[
  \begin{tikzcd}
  \hPCSP(A,B)
      \arrow[d, "A\circ\gr(\ARG)"']
  &&
  \hPCSP(A',B')
  \\
  \sPCSP(\id,\Ran_{A} B)
      \arrow[rr, "\text{do nothing}"']
  &&
  \sPCSP(\id,\Ran_{A'} B')
      \arrow[u, "\gl{\ARG}\circ A'"']
  \end{tikzcd}
\]
The reductions on either side are given by Theorem~\ref{t:fundamental-theorem}, and the ``doing nothing'' map is a reduction since $\Ran_{A'} B'\to \Ran_{A} B$. Moreover, since $\cat S$ is finite, the sizes of the sets $A(s)$, for $s\in\cat S$, is bounded from above by some constant $k$, hence the same holds for the functors of the form $A\circ \gr(D)$. Consequently, the composition of the reductions is computable in log-space.
\end{proof}

As another corollary, we obtain a hardness criterion sufficient to prove NP-hardness side of the Bulatov--Zhuk dichotomy. This criterion was first proved for finite-template CSPs in \cite[Corollary 7.3]{BulatovJK05}, and for PCSPs in \cite[Corollary 5.2]{BBKO21}.

\begin{corollary} \label{cor:bulatov-zhuk-NP-complete}
  If there is a natural transformation $\Ran_A B\to \id$, then $\hPCSP(A,B)$ is \NP-complete.
\end{corollary}

\begin{proof}
    We prove the \NP-hardness by a reduction from graph $3$-colouring, i.e.\ the problem $\hCSP(K_3)$; see Example~\ref{ex:3-col}.
    First, observe that $\id \to \Ran_{K_3} K_3$ (by \eqref{eq:ran} applied to the identity $K_3 \to K_3$). Consequently, we have a natural transformation $\Ran_A B \to \Ran_{K_3} K_3$ which induces a log-space reduction from $\hCSP(K_3)$ to $\hPCSP(A, B)$ by Theorem~\ref{thm:ranGivesReductions}. Since the former problem is \NP-hard, so is the latter.
\end{proof}

Note that in the above proof, we implicitly proved that $\sCSP(\id)$ is \NP-complete on instances involving sets of size at most $6$. In fact, the problem is NP-hard on instances with sets of size at most~$3$ which can be proved by a reduction from 1-in-3-SAT (whose hardness follows from the above corollary as well).

The above corollary is enough to prove all NP-hardness within finite-template CSPs (unless P = NP); see Theorem~\ref{thm:bulatov-zhuk}. Nevertheless, there are finite-template PCSPs that do not satisfy the assumption of the corollary, but whose hardness is proved using Theorem~\ref{thm:ranGivesReductions}.

\begin{example}
  Using algebraic topology, Avvakumov et al.~\cite{AvvakumovFOTW25} prove that the problem $\hPCSP(C_{2k+1}, K_4)$ is NP-complete for an undirected odd cycle $C_{2k+1}$ of length $2k+1$, where $k \geq 1$, and a 4-clique $K_4$.
  The proof follows the line established by Krokhin et al.~\cite[Section 3]{KOWZ23}: We first assign to each graph a simplicial complex (a topological space) and then analyse the polymorphisms of the graphs using the topological structure. The final step is then a combination of Theorem~\ref{thm:ranGivesReductions} and a known hardness criterion \cite[Proposition 5.14]{BBKO21}.
  More concretely, we obtain two natural transformations%
  \footnote{The map $\mu$ described in \cite[Sec.~1, Step~1]{AvvakumovFOTW25} is not natural itself, but can be made natural by replacing $S^1$ and $S^2$ with homotopically equivalent simplicial sets.}
  \[\begin{tikzcd}
    \pol(C_{2k+1}, K_4) \arrow[r, "\mu"]  &
    \pol(S^1, S^2)      \arrow[r, "\eta"] &
    B
  \end{tikzcd}\]
  where $S^1$ and $S^2$ are triangulations of the 1-dimensional circle and the 2-dimensional sphere (equipped with an action of $\mathbb Z_2$) and $B\colon \fin \to \fin$ is such that $\sPCSP(\id, B)$ satisfies the hardness criterion given in \cite[Proposition 5.14]{BBKO21}.
  Consequently, the hardness of $\hPCSP(C_{2k+1}, K_4)$ follows from Theorem~\ref{thm:ranGivesReductions}.

 Although this might appear as a two step reduction with an intermediate problem $\sPCSP(S^1, S^2)$, Avvakumov et al.~\cite{AvvakumovFOTW25} do not discuss the complexity of the intermediate problem.
  The reason for this is that simplicial complexes are not relational structures and hence \cite{BBKO21} does not apply to this case.
  Naturally, you can find an ad-hoc combinatorial argument to show that indeed there is such a reduction, but that requires additional work.
  In contrast, our approach immediately applies to simplicial sets, which can be used in place of simplicial complexes, since they are presheaves on the (locally finite) category $\Delta$ of finite ordinals and non-decreasing maps. We can also encode the $\mathbb Z_2$ action by adding some arrows to~$\Delta$. Consequently, we derive that (the simplicial version of) $\sPCSP(S^1, S^2)$ is NP-hard for any triangulations of $S^1$ and $S^2$.
\end{example}
\subsection{Minor conditions}
    \label{sec:minor-conditions}

Barto et al.~\cite[Theorem 3.12]{BBKO21} formulate Theorem~\ref{t:fundamental-theorem} using a problem very similar to our $\sPCSP(\id_\fin, M)$ denoted by $\operatorname{PMC}_n(M)$. The key difference is that inputs to the problem $\operatorname{PMC}_n(M)$ are interpreted as \emph{minor conditions} rather than functors $\cat J \to \fin$. In this section, we explain the relation between such minor conditions and diagrams, and compare the constructions used in \cite{BBKO21} with constructions used in this paper.

We start by introducing minor conditions, which are a special case of a general algebraic notion of \emph{Maltsev conditions}.

\begin{definition}
  A \emph{minor identity} is a formal expression 
  \[
    f(x_{\pi(1)}, \dots, x_{\pi(m)}) \equals g(x_1, \dots, x_n)
  \]
  for some $\pi\colon [m] \to [n]$, where $f$ and $g$ are function symbols, and $x_1, \dots, x_n$ are distinct variables. The symbol $\equals$ is used to distinguish such formal identities from equality of terms. Importantly, the function symbols, e.g., $f$ in the example, are treated as placeholder symbols not as actual functions.

  A \emph{minor condition} is a finite set $\Gamma$ of such minor identities.
  We say that a minor condition \(\Gamma\) is satisfied in \(\pol(\mathbf A, \mathbf B)\) if there exists and interpretation \(\xi\) of function symbols occurring in \(\Gamma\) as polymorphisms \(\mathbf A^n \to \mathbf B\) such that all identities \(f \equals g\) in \(\Gamma\) are true, i.e.\ \(\xi(f)(a_{\pi(1)}, \dots, a_{\pi(m)}) = \xi(g)(a_1, \dots, a_n)\) for all \(a_1,\dots,a_n\) in \(\mathbf A\).
\end{definition}

\begin{example} \label{ex:siggers-1}
  The Siggers identity \cite{Siggers10}, which is often used to delineate the boundary between tractable and NP-complete finite-template CSPs, is usually written as
  \[
    s(x, y, z, x, y, z) \equals 
    s(y, z, x, z, x, y).
  \]
  Although this is not strictly speaking a minor identity, we can rewrite it to an equivalent minor condition consisting of two identities:
  \begin{align}
    s(x, y, z, x, y, z) \equals t(x, y, z) \label{eq:sg1}\\
    s(y, z, x, z, x, y) \equals t(x, y, z) \label{eq:sg2}
  \end{align}

  By definition, given a structure $\mathbf A$, $\pol(\mathbf A)$ satisfies the Siggers condition if and only if there is a homomorphism $s\colon \mathbf A^6 \to \mathbf A$ such that
  \[
    s(a_1, a_2, a_3, a_1, a_2, a_3) =
    s(a_2, a_3, a_1, a_3, a_1, a_2)
  \]
  for all $a_1, a_2, a_3 \in D^\mathbf A$.
  In particular, it is not hard to observe that $\pol(K_3, K_k)$, for any $k\geq 3$, does not satisfy the Siggers identity. This is since $s\colon [3]^6 \to [k]$ would need to satisfy $s(0, 1, 2, 0, 1, 2) = s(1, 2, 0, 2, 0, 1)$ which contradicts preservation of edges.
\end{example}

Next we observe that satisfaction of a minor condition \(\Gamma\) is the same as admitting a solution to a functor $D_{\,\Gamma}\colon \cat J_{\,\Gamma} \to \fin$ built from~\(\Gamma\).
Objects of the category $\cat J_{\,\Gamma}$ are the function symbols occurring in~$\Gamma$ and we include a morphism $e_\pi\colon f \to g$ for each identity in $\Gamma$ of the form $f(x_{\pi(1)}, \dots, x_{\pi(m)}) \equals g(x_1, \dots, x_m)$.
The functor $D_{\,\Gamma}$ then maps an object $f$ to the set $[n]$ where \(n\) is the arity of \(f\), and a morphism $e_\pi$ to $\pi$.
Similarly to Remark~\ref{ex:pp}, observe that indeed $\pol(\mathbf A, \mathbf B)$ satisfies $\Gamma$ if and only if $\pol(\mathbf A, \mathbf B) \circ D_{\,\Gamma}$ has a solution.

\begin{example}[Example \ref{ex:siggers-1}, cont.] \label{ex:siggers-2}
    To conclude our example, we describe the functor $D_{\,\Gamma}$ for the Siggers condition $\Gamma$. The category~$\cat J_{\,\Gamma}$ has two objects $s, t$ and two morphisms $\alpha_0, \alpha_1 \colon s \to t$, each corresponding to the identities \eqref{eq:sg1} and \eqref{eq:sg2}. Therefore, $D_{\,\Gamma}(s) = [6]$ and $D_{\,\Gamma}(t) = [3]$ and \(D_{\,\Gamma}(\alpha_i) = \pi_i\colon [6] \to [3]\) are as specified in \eqref{eq:sg1} and \eqref{eq:sg2}.
    Observe that the signature category $\cat D$ of digraphs (from Example~\ref{ex:graphs-as-presheaves}) is equivalent to $\cat J_{\,\Gamma}$. Therefore, $D_{\,\Gamma}$ can be interpreted as a graph. In fact, \(D_{\,\Gamma}\) is isomorphic to the copresheaf $K_3$ from Example~\ref{ex:3-col}.
    The conditions corresponding to the diagrams $\mathbf G^\mkPrsh$ given by a graph $\mathbf G$ are called \emph{loop conditions} \cite[Definition 1]{Olsak19}.
\end{example}

It is immediate to see that every functor \(\cat J \to \fin\), where \(\cat J\) is finite, can be obtained as \(D_{\,\Gamma}\) for some minor condition \(\Gamma\).
Therefore, $\PMC_n(M)$ from \cite[Definition 3.8]{BBKO21} is equivalent to the restriction of the problem $\sPCSP(\id, M)$ to the instances whose image contains sets of size at most $n$.
Moreover, given relational structures $\mathbf A, \mathbf B, \mathbf I$, the two constructions $\Sigma(\mathbf A, \mathbf I)$ and $\mathbf I_\Gamma(\mathbf A)$ used in \cite[Sections 3.2 and 3.3, resp.]{BBKO21} to reduce between $\hPCSP(\mathbf A, \mathbf B)$ and $\PMC_n(\pol(\mathbf A, \mathbf B))$ correspond precisely to our reductions in the proof of Theorem~\ref{t:fundamental-theorem}:
\begin{align*}
  D_{\Sigma(\mathbf A, \mathbf I)}      &\simeq \mathbf A^\mkPrsh \circ \gr \mathbf I^\mkPrsh \\
  \mathbf I_\Gamma(\mathbf A)           &\simeq (\gl {D_{\,\Gamma}} \circ \mathbf A^\mkPrsh)^\mkStr
\end{align*}

\begin{remark}\label{rem:morphisms-preserving-MCs}
    We note that the traditional notion of homomorphisms between minor conditions, called \emph{interpretations}, do not align with homomorphisms in the category $\fcat \comma \fin$.
    Instead, they align with the Kleisli morphisms of the monad $\gr\gl\ARG$ arising from the adjunction $\gl\ARG \dashv \gr$.

    Interpretations have the following semantics: $\Pi$~is \emph{interpretable in} $\Gamma$ if and only if whenever a (function) minion $M$ satisfies $\Gamma$ then $M$ satisfies $\Pi$.
    Note that the Kleisli morphisms of $\gr\gl\ARG$ are functors \(H\) from the domain of \(D_\Gamma\) to that of \(\gr \gl{D_\Pi}\) such that \(D_\Gamma = \gr \gl{D_\Pi} \circ H\).
    Recall, by Lemma~\ref{l:gl-reduction}, that a minion \(M\) satisfies \(\Pi\) if and only if $\gl {D_\Pi} \to M$.
    Therefore, \(D_\Gamma \to \gr \gl{D_\Pi}\), which is equivalent to \(\gl{D_\Gamma} \to \gl{D_\Pi}\), gives that, if $\gl {D_\Pi} \to M$, then also $\gl {D_\Gamma} \to M$, i.e.\ $\Pi$~is interpretable in $\Gamma$.
    Conversely, if $\Pi$ is interpretable in $\Gamma$, since \(\gl{D_\Pi} \to \gl{D_\Pi}\), it follows that \(\gl{D_\Pi} \to \gl{D_\Gamma}\) or, equivalently, there is a Kleisli morphism \(D_\Pi \to \gr\gl{D_\Gamma}\).
\end{remark}

\section{Classifying gadget reductions}
  \label{s:adjunctions} \label{sec:gadget-reductions} \label{sec:gadgets}

Theorem~\ref{thm:ranGivesReductions} gives a sufficient condition for the existence of a log-space reduction between two PCSPs.
Such result is enough for the study of computational complexity of PCSPs.
Nevertheless, this statement hides another strength of the algebraic approach, namely that reductions arising in Theorem~\ref{thm:ranGivesReductions} correspond precisely to a special type of log-space reductions.

These reductions are given by \emph{gadget constructions}, which is a common technique in finite model theory and combinatorics; see, e.g.,~\cite{HellN90}.
They have a slightly stricter definition in the theory of CSPs, which we present in Definition~\ref{def:gadgets} below.
In Section~\ref{sec:5.5-proof2}, we make use of our categorical reformulation to give a sleek proof of the \emph{fundamental theorem} of the algebraic approach to CSP (which incidentally provides an alternative proof of Theorem~\ref{t:fundamental-theorem}).

\begin{theorem}
  \label{thm:gadgets}
 Consider promise templates \((A,B)\) and \((A',B')\) where $A, B \colon \cat S \to \fin$ and $A', B'\colon \cat T \to \fin$ are finite copresheaves.
    Then, the following are equivalent:
  \begin{enumerate}
      \item \(\hPCSP(A,B)\) reduces to \(\hPCSP(A',B')\) via a gadget reduction
        (and hence in log-space);
      \item there is a natural transformation $\Ran_{A'} B' \to \Ran_A B$.
  \end{enumerate}
\end{theorem}

\subsection{Adjoints provide reductions}

We start with the observation that right adjoints may be used for reductions between two (P)CSPs.
The following observation can be attributed to \cite[Observation 2.10]{WrochnaZ20} and \cite[Theorem 4.4]{KOWZ23}, although a similar observation was already made in \cite[pp.~93--94]{HellN90}.

\begin{lemma} \label{l:marcins}
    Let \(L\colon \cat C \to \cat C'\) and \(R\colon \cat C' \to \cat C\) be a pair of adjoint functors where \(L\dashv R\), i.e., $L$ is the left adjoint to $R$. Now consider the two homomorphism problems $\hPCSP(A,B)$ and $\hPCSP(A',B')$ for $A,B\in\cat C$ and $A',B'\in\cat C'$.

     Then \(L\) is a reduction from \(\hPCSP(A,B)\) to \(\hPCSP(A',B')\) if and only if \(A \to R(A')\) and \(R(B') \to B\).
\end{lemma}
\begin{proof}
    First, assume that $L$ is a reduction. Since \(A \to A\), preservation of \Yes{} instances by \(L\) implies \(L(A) \to A'\), which is equivalent to \(A \to R(A')\). Similarly, the adjunction provides a morphism \mbox{\(LR(B') \to B'\)}, implying that $LR(B')$ is not a \No{}  instance. But since $L$ preserves \No{} instances, we must have \(R(B') \to B\).

     Conversely, take an instance $X\in\cat C$. To show that $L$ is a reduction, note that if \(X \to A\) then \(L(X) \to L(A) \to A'\) and if \(L(X) \to B'\) then also \(X \to R(B') \to B\).
\end{proof}

It should be noted that a weaker condition is enough to derive a reduction: It is enough that \(L(A) \to A'\) if and only if \(A \to R(A')\), i.e., the two hom sets do not need to be isomorphic.
In other words, it is enough to have an adjunction between the posetification of the two categories.
While reductions arising from proper adjunctions are fully characterised by the theory presented in this section, characterising reductions arising from posetal adjunctions is an important open problem in PCSPs (see, e.g., \cite[Section~4]{KrokhinO22}).

\subsection{Yoneda extensions}
\label{s:gadget-reductions}

Particularly relevant for (P)CSP reductions are the adjunctions studied already by Kan~\cite[Remark on p.~333]{Kan58}; the Yoneda extension and generalised nerve adjunctions.
We start by describing the former functors, which are the left adjoints (cf. \cite{Riehl2016context}~Remark 6.5.9).

\begin{definition}
    Let \(G\colon \cat S^\op \to \cat C\) be a functor from a finite category \(\cat S\) to a locally finite and finitely cocomplete category $\cat C$.
  The \emph{Yoneda extension} of $G$ is the functor $\kay G \colon \pshcat S \to \cat C$ defined by
  \begin{equation}
    \kay G(A) = \colim(G\circ (\gr A)^\op) %
    \label{eq:kay}
  \end{equation}
  for all $A \colon \cat S \to \fin$.
\end{definition}

We note that the Yoneda extension of a functor $G$ also exists in cases where $\cat S$ is not finite, as long as $\cat C$ admits the colimit in~\eqref{eq:kay} for every $A\colon \cat S \to \fin$.
In our typical setting, \(\cat C\)~is a concrete category such as another category of copresheaves, graphs or structures.
Then, the computation of \(\kay G(A)\), using the above colimit formula, corresponds well to the intuitive notion of gadget construction. Namely, \(\kay G(A)\) is obtained by replacing each element of a copresheaf $A\colon \cat S \to \fin$, i.e., an object $(s, a)$ of $\el A$, with the \emph{gadget} $G(s) \in \cat C$ and by `glueing' these gadgets according to the image of morphisms under $G$.
We demonstrate this in the following example.

\begin{example}[Example~\ref{ex:graphs-as-presheaves} cont.]
    \label{ex:realis-concretely}
    Let us consider the case where \(\cat S\) is the signature category \(\cat D\) of digraphs and \(\cat C\) is the category of graphs $\graphs$.
    To specify a functor $G\colon \cat D^\op \to \graphs$ it is enough to select a pair of graphs $G(V)$ and $G(E)$ and a pair of homomorphisms
    \[
      G(s), G(t)\colon G(V) \to G(E).
    \]
    Given such a functor and a multidigraph $A \colon \cat D \to \fin$, the graph $\kay G(A)$ is constructed by replacing each vertex of $A$ by a copy of $G(D)$ and each edge $e\in A(E)$ by a copy of $G(E)$. These copies are `glued' together according to the maps~$G(s)$ and~$G(t)$, as prescribed by the colimit computation (cf. Example~\ref{ex:gl} and Figure~\ref{fig:gl}).

    More concretely, let $G(V)$ be the graph with a single vertex $\star$ and $G(E)$ a path of length~$3$, i.e., the graph
    \[
    \begin{tikzpicture}
      [every node/.style = {circle, fill, inner sep = 0, outer sep = 1pt, minimum size = 4pt}]
      \node (0) [label=left:$0$] {};
      \node (1) at (1, 0) {};
      \node (2) at (2, 0) {};
      \node (3) [label=right:$1$] at (3, 0) {};
      \draw (0) -- (1); \draw (1) -- (2); \draw (2) -- (3);
    \end{tikzpicture}
    \]
    and let $G(s) \colon {\star} \mapsto {0}$ and $G(t) \colon {\star} \mapsto {1}$.
    The graph $\kay G(A)$ is obtained from $A$ by subdividing each edge into three edges, e.g., if $A$ has two vertices connected by two parallel edges, then $\kay G(A)$ is a $6$-cycle.
\end{example}

\subsection{Generalised nerves}

We recall a well-known fact that, under mild assumptions, Yoneda extensions admit right adjoints, which we call \emph{generalised nerves}. Later we use this fact to employ Lemma~\ref{l:marcins} when checking that gadget constructions are reductions.

\begin{definition}
    Given a functor \(G\colon \cat S^\op \to \cat C\) from a finite category $\cat S$ to a locally finite category $\cat C$.
    The \emph{(generalised) nerve of~\(G\)} is the functor
\(
    \nerve G \colon \cat C \to \pshcat S
\) defined by
\[
  \nerve G(B)(s) = \hom_{\cat C}(G(s), B).
\]
for all objects $B\in\cat C$ and $s\in\cat S$
\end{definition}

\begin{example}[Example~\ref{ex:realis-concretely} cont.]
    \label{ex:nerve-concretely}
  Let \(G \colon \cat D \to \graphs\) be the functor from Example~\ref{ex:realis-concretely}.
  For any graph \(A\in \graphs\), the multidigraph \(B=\nerve G(A)\in\pshcat D\) is then computed as follows:
  \begin{itemize}
    \item The set of vertices \(B(V)\) of \(B\) is given by \(\hom(G(V),A)= \hom(\star, A)\) which is the set of vertices of $A$.
    \item The set of edges is given by \(B(E) = \hom(G(E), A)\). For two fixed vertices $v,w\in B(V)$ the edges in $e\in B(E)$ from $v$ to $w$ (i.e.\ $s(e)=v$ and $t(e)=w$) precisely correspond to morphisms $f\colon G(E)\to A$ with $f(0) = v$ and $f(1) = w$. Since the graph $B(E)$ is a path of length $3$, such $f$ corresponds to the walks in $A$ of length $3$ from $v$ to $w$.
  \end{itemize}
  For example, $\nerve G(\mathbf C_5)$, where $\mathbf C_5$ is a $5$-cycle, is up to multiplicity of edges the copresheaf corresponding to the complete graph $\mathbf K_5$ with $5$ vertices.
\end{example}

\begin{figure}[t]
    \[
  \begin{tikzcd}[row sep=1em]
  & {\cat C} \arrow[dd, "\nerve G", bend left=20, xshift=0.2em]
  \\
  {\cat S^\op}
    \arrow[ru, "G"]
    \arrow[rd, swap, hook, "\yo"]
  \\
  & {\pshcat S}
    \arrow[uu, "\kay G", bend left=20]
    \ar[phantom,xshift=0.1em]{uu}{\dashv}
  \end{tikzcd}
\]
    \caption{The setting of Propositions~\ref{p:nerve-realization} and~\ref{p:nerve-realization-unique}.}
    \label{fig:nerve-realization}
\end{figure}

\begin{proposition}
  \label{p:nerve-realization}
  Let \(G\colon \cat S^\op \to \cat C\) be a functor from a finite category $\cat S$ to a locally finite and finitely cocomplete category $\cat C$.
  Then \(\kay G\) is the left adjoint of \(\nerve G\), i.e.\ $\kay G\dashv \nerve G$ as in Fig.~\ref{fig:nerve-realization}.
\end{proposition}

\begin{proof}
  We need to show that there is a natural equivalence $\hom(\kay G(A), C) \simeq \hom(A, \nerve G(C))$ where $A\colon \cat S \to \fin$ and $C \in \cat C$. Substituting the definitions of $\kay G(A)$ and $\nerve G(C)$, this follows from the following computation:
  \begin{multline*}
    \hom(\kay G(A), C)
       = \hom \bigl(\colim(G\circ (\gr A)^\op), C \bigr) \\
       \simeq \lim \bigl(\hom({-},C)\circ G^\op\circ\gr A \bigr)
       \simeq \hom \bigl(A, \hom({-},C)\circ G^\op\bigr)
       = \hom(A, \nerve G(C) )
  \end{multline*}
  where the first equivalence follows since the contravariant functor $\hom({-}, C)$ maps colimits to limits, and the second equivalence is Lemma~\ref{lem:gr1} applied to $A \coloneq \hom({-}, C)\circ G^\op$ and $X \coloneq A$.
\end{proof}

\begin{proposition}
    \label{p:nerve-realization-unique}
    Let $\cat S, \cat C$ and $\yo$ be as in Proposition~\ref{p:nerve-realization}.
    Every adjoint pair $L\dashv R$, where $L$ is a functor $\pshcat S\to \cat C$, is naturally isomorphic to $\kay{G}\dashv\nerve G$ for $G= L\circ \yo$.
\end{proposition}

\begin{proof}
    Observe that, by the Yoneda lemma, $\nerve{\yo} \cong\id_{\pshcat S}$, i.e.\ the nerve $\nerve{\yo}$ is isomorphic to the identity functor on ${\pshcat S}$. Then Proposition~\ref{p:nerve-realization} implies that $\kayyo\cong\id_{\pshcat S}$ which implies the so called density formula for copresheaves~\cite[Theorem 6.5.7]{Riehl2016context}:
    \[
      \colim(\yo\circ(\gr A)^\op) = \kayyo(A)\cong A.
    \]
    In particular, every copresheaf is a colimit of copresheaves in the image of $\yo$.
    Using this and the fact that left adjoint functors preserve colimits, one computes
    \[
      \kay{L\circ \yo}(A) =
      \colim( L\circ \yo \circ (\gr A)^\op )
      \simeq
      L\bigl( \colim (\yo \circ (\gr A)^\op )\bigr) \simeq L(A).
      \qedhere
    \]
\end{proof}

\begin{remark}
The fact that gadget constructions admit right adjoints was first observed in the CSP context in \cite[pp.~93--94]{HellN90}.
The right adjoints appear more prominently in the CSP literature; they correspond to the notion of \emph{pp-interpretations} (see Appendix~\ref{sec:pp-interpretations}).
\end{remark}

\subsection{Categorical gadget reductions}

We define gadget reductions between PCSPs as those reductions which are given by a gadget construction. %

\begin{definition}[Gadget reduction] \label{def:gadgets}
  Let $(A,B)$ and $(A', B')$ be promise templates such that $A, B \colon \cat S \to \fin$ and $A', B' \colon \cat T \to \fin$ are finite copresheaves.
  A \emph{gadget reduction} from $\hPCSP(A, B)$ to $\hPCSP(A', B')$ is a functor
  \[
    \kay G\colon \pshcat S \to \pshcat T
  \]
  for some $G\colon \cat S \to \pshcat T$ which is a valid reduction, i.e., such that, for all $X \colon \cat S \to \fin$,
  \begin{description}
    \item[Soundness:] if $X \to A$, then $\kay G(X) \to A'$; and
    \item[Completeness:] if $\kay G(X) \to B'$, then $X \to B$.
  \end{description}
\end{definition}

Importantly, if both $\cat S$ and $\cat T$ are finite categories, the functor $G\colon \cat S^\op \to \pshcat T$ is given by a finite amount of data, and we can compute the copresheaf $\kay G(A)$ in log-space. This is again a consequence of Lemma~\ref{lem:reingold}.

\begin{lemma}
  Let $\cat S$ and $\cat T$ be finite categories, and let $G\colon \cat S^\op \to \pshcat T$ be a functor. Then the functor $A \mapsto \kay G(A)$ is computable in log-space.
\end{lemma}

It follows directly from Proposition~\ref{p:nerve-realization-unique} that gadget reductions between PCSPs are the same as left adjoints validating the conditions of Lemma~\ref{l:marcins}.

\begin{proposition}
    \label{p:gadgetred-as-adjoints}
    Let \((A,B)\) and \((A',B')\) be promise templates where $A, B\colon \cat S \to \fin$ and $A', B'\colon \cat T \to \fin$ are finite copresheaves.
    Then, the following are equivalent:
  \begin{enumerate}
      \item \(\hPCSP(A,B)\) reduces to \(\hPCSP(A',B')\) via a gadget reduction (and hence in log-space);
      \item \(\hPCSP(A,B)\) reduces to \(\hPCSP(A',B')\) via a left adjoint;
      \item there is a functor $G\colon \cat S^\op \to \pshcat T$ and morphisms $A\to\nerve G A'$ and $\nerve G B'\to B$; and
      \item there is a right-adjoint functor $R\colon\pshcat T\to\pshcat S$ and morphisms $A\to RA'$ and $RB'\to B$.
  \end{enumerate}
\end{proposition}

\begin{proof}
    Recall that, by Proposition~\ref{p:nerve-realization}, adjunctions between copresheaf categories are always of the form \mbox{$\kay G \dashv \nerve G$} for some $G$.
    This shows the equivalence of (1) and (2), and of (3) and (4).
    Finally, by Lemma~\ref{l:marcins}, a left adjoint $L$ is a reduction, i.e.\ (2) holds, if and only if its right adjoint $R$ satisfies the conditions of (4).
\end{proof}

\begin{remark}
  In the CSP and PCSP literature, the expression involving nerve in item (3) of the above proposition would be phrased as ``$A, B$ is \emph{pp-constructible} from $A', B'$'', see Appendix~\ref{sec:pp-interpretations}.
\end{remark}

\subsection{Kan strikes back}
  \label{sec:5.5-proof2}

In this section, we prove Theorem~\ref{thm:gadgets}.
The implication (2)$\to$(1) decomposes into two reductions which go via the intermediate problem of the form $\hPCSP(\id_\fin, M)$ where $M \colon \fin \to \fin$ is a polymorphism minion.

\begin{theorem} \label{thm:3.12}
  Let $A, B\colon \cat S \to \fin$ be finite copresheaves. Then
  $\hPCSP(A,B)$ and $\hPCSP(\id_\fin, \Ran_A B)$ are equivalent up to gadget reductions.
\end{theorem}

\begin{proof}
  The translation from left to right is given by the left Kan extension $\Lan_A$\,, which exists since $\cat S$ is finite and $[\fin, \fin]$ has finite colimits~\cite[Theorem 6.2.1]{Riehl2016context}.
  Using Lemma~\ref{l:marcins} we see that $\Lan_A$\, is a reduction from $\hPCSP(A, B)$ to $\hPCSP(\id, \Ran_A B)$ since, for its right adjoint \((\ARG\circ A)\), we have the identity morphism \(A \to \id\circ A\) and also a morphism $(\Ran_A B) \circ A \to B$ as the counit of the adjunction $({\ARG}\circ A) \dashv \Ran_A$\,.

  The other reduction, which is $({\ARG}\circ A)$, is provided by Lemma~\ref{l:precomp-reduction}. Both are gadget reductions by Proposition~\ref{p:gadgetred-as-adjoints} and the fact that they are left adjoints.
\end{proof}

The implication (2)$\to$(1) of Theorem~\ref{thm:gadgets} is now a corollary of Theorem~\ref{thm:3.12}.
To this end, assume that $\Ran_{A'} B' \to \Ran_A B$ and consider the following diagram.
\[
\begin{tikzcd}[row sep=scriptsize]
\hPCSP(A,B)
    \arrow[d, "\Lan_A"']
&&
\hPCSP(A',B')
\\
\hPCSP(\id,\Ran_{A} B)
    \arrow[rr, "\text{do nothing}"']
&&
\hPCSP(\id,\Ran_{A'} B')
    \arrow[u, "(\ARG\circ A)"']
\end{tikzcd}
\]
The vertical arrows are reductions by Theorem~\ref{thm:3.12} and the horizontal (doing nothing) is a reduction since $\Ran_{A'} B' \to \Ran_A B$.
Consequently, the composite reduction is also a left adjoint, and hence it is a gadget reduction by Proposition~\ref{p:gadgetred-as-adjoints}.
For the other implication (1)$\to$(2), we use an observation about product preserving functors (in particular about right adjoints) and their interactions with polymorphism minions; cf.~\cite[Lemma 4.8]{WrochnaZ20} and \cite[Lemma~C.15]{FilakovskyNOTW24}.

\begin{lemma} \label{l:r-minionHom}
  Let $R\colon \cat C\to \cat C'$ be a finite products preserving functor, \(\cat C, \cat C'\) locally finite, and let $A, B\in \cat C$.
  Then $R$ induces a natural transformation $\pol(A,B)\to \pol(RA, RB)$.
\end{lemma}

\begin{proof}
  Since \(\pol(A,B)(N) = \hom(A^N, B)\), functoriality of \(R\) induces a natural transformation $\hom(A^N, B) \to \hom(R(A^N), RB)$.
  Since $R$ preserves finite products, the latter is naturally equivalent to
  \[
    \hom(R(A^N), RB )\cong \hom((RA)^N, RB) = \pol(RA, RB)(N). \qedhere
  \]
\end{proof}

The implication (1)$\to$(2) of Theorem~\ref{thm:gadgets} is derived as follows:
Since $\Ran_A B \cong \pol(A, B)$ by Lemma~\ref{lem:pol-is-ran}, it is enough to construct some natural transformation $\pol(A', B') \to \pol(A, B)$.
From a gadget reduction $L$ from $\hPCSP(A, B)$ to $\hPCSP(A', B')$, we obtain, by Proposition~\ref{p:gadgetred-as-adjoints}, a right adjoint $R$ such that $A \to RB$ and $RB' \to A'$. Consequently, we have
\[
  \pol(A', B') \to \pol(RA', RB') \to \pol(A, B)
\]
where the first natural transformation is given by Lemma~\ref{l:r-minionHom}, and the second follows from $A \to RA'$, $RB' \to B$, and functoriality of~$\pol(\ARG,\ARG)$.
This concludes the proof of Theorem~\ref{thm:gadgets}.

\begin{remark} As a consequence of the proof, we get that if $\hPCSP(A, B)$ reduces to $\hPCSP(A', B')$ via a gadget reduction, then $X\mapsto (\Lan_A X)\circ A'$ is a reduction as well. In other words, there is a \emph{universal gadget reduction}.
\end{remark}

\subsection{A formula for the left Kan extension}

Given a natural transformation $\Ran_{A'} B' \to \Ran_A B$, Theorems~\ref{thm:ranGivesReductions} and~\ref{thm:gadgets} induce two possible reductions from $\hPCSP(A, B)$ to $\hPCSP(A', B')$. The former reduction maps a copresheaf \(X\) to \(\gl{A\circ \gr X} \circ A'\) and the latter to \((\Lan_A X) \circ A'\).

The following proposition shows that these reductions coincide which, as a by-product, entails that the reduction given by Theorem~\ref{thm:gadgets} is a gadget reduction.

\begin{proposition}
  \label{weirdformulaforlan}
  Let $A,X \colon \cat S \to \fin$ be finite copresheaves, then
  \[
  \Lan_A X \cong \gl{A\circ \gr X}.
  \]
\end{proposition}

\begin{proof}
  We observe that $X\mapsto \gl{A\circ \gr X}$ is a left adjoint to $M \mapsto M\circ A$, since
  \[
    \hom (\gl{A\circ \gr X}, M) \simeq
    \lim (M \circ A \circ \gr X) \simeq
    \hom (X, M\circ A)
  \]
  where the first equivalence follows by Lemma~\ref{l:gl-reduction} and the second by Lemma~\ref{lem:gr1}.
  The rest follows from the uniqueness of adjoints.
\end{proof}

\section{Bulatov--Zhuk theorem for copresheaves}
    \label{sec:bulatov-zhuk}

In this subsection, we prove the extension of the Bulatov--Zhuk dichotomy to all CSPs whose template is a finite copresheaf, i.e.\ Theorem~\ref{thm:bulatov-zhuk-for-presheaves}.
We derive this theorem from the dichotomy for finite structures, which was proved by Bulatov \cite{Bulatov17} and Zhuk \cite{Zhuk20}.

\begin{theorem}[Bulatov--Zhuk] \label{thm:bulatov-zhuk}
  Let $\mathbf A$ be a finite relational structure such that there is no natural transformation $\pol(\mathbf A) \to \id$, then $\hCSP(\mathbf A)$ is in \Ptime.
\end{theorem}

Note that Theorem~\ref{thm:bulatov-zhuk-for-presheaves} generalises the Bulatov--Zhuk theorem in the following way:
The transformations $\mathbf X \mapsto \mathbf X^\vartriangle$ and $Y \mapsto Y^\triangledown$ provide log-space reductions between $\hCSP(\mathbf A)$ and $\hCSP(\mathbf A^\vartriangle)$.
Moreover, the theorem immediately applies to structures with unary functional symbols. Note that Feder, Madelaine, and Stewart~\cite{FederMS04} showed that the homomorphism problem of structures with two unary symbols has as rich complexity classification as finite-template CSPs (up to polynomial-time Turing reductions).
Our reformulation of Bulatov--Zhuk not only shows a \Ptime/\NP-complete dichotomy of these problems, but also that the boundary aligns with the well-established algebraic boundary for (relational) CSPs.

Theorem~\ref{thm:bulatov-zhuk-for-presheaves} is a direct corollary of the following proposition, by a reduction to the case of single-sorted relational structures.
In the proof, we use a standard encoding of multi-sorted structures as single-sorted; see, e.g.\ \cite[Definition~5 and Proposition~1]{BulatovJ03}.

\begin{definition} \label{def:single-sorted-from-presheaf}
Given a copresheaf $A\colon \cat S \to \fin$, we construct a relational structure $\mathbf A'$ in the signature with a binary relation $E_\pi$ for each morphism $\pi \colon s \to t$ in $\cat S$ (including the identity morphisms):
The domain of~$\mathbf A'$ is $\prod_{s\in \cat S} A(s)$.
The relation $E_\pi^{\mathbf A'}$, for a morphism $\pi \colon s \to t$, is defined as
$
  E_\pi^{\mathbf A'} =
    \{ (a, b) \mid A(\pi)(a_s) = b_t \}
$.
\end{definition}

\begin{proposition} \label{prop:presheaves-to-structures}
  For finite copresheaves $A, B \colon \cat S \to \fin$, and the relational structures $\mathbf A'$ and $\mathbf B'$ constructed from $A$ and $B$, respectively,
  \[
    \pol(\mathbf A', \mathbf B') \simeq \Ran_A B.
  \]
\end{proposition}

\begin{proof}
  The natural isomorphism $\pol(\mathbf A', \mathbf B') \simeq \Ran_A B$ is constructed as follows.
  We define a mapping
  \[
    \xi_N \colon \hom(A^N, B) \to \hom((\mathbf A')^N, \mathbf B')
  \]
  by sending $f\colon A^N \to B$ to the map $g\colon \bigl(\prod_{s\in \cat S} A(s)\bigr)^N \to \prod_{s\in \cat S} B(s)$ that applies $f$ coordinate-wise. More precisely, the $s$-th component of the result is obtained by applying $f_s$ to the $N$-tuple of $s$-th components of the inputs.
  It is not hard to check that this mapping is natural, well-defined, and each $\xi_N$ is injective.

  To show that $\xi_N$ is an equivalence, it is enough to observe that it is surjective.
  Assume that $g \in \hom((\mathbf A')^N, \mathbf B')$.
  First, we show that $g$ acts component-wise on tuples in $\mathbf A'$, i.e., that the $s$-th components of $g(b)$ is determined by the $s$-th components of $b(i)$'s where $i\in N$. This follows from the fact that $g$ preserves the relation $E_{\id_s}$ where $\id_s$ is the identity morphism on $s$. Indeed, if $b, b' \in (D^{\mathbf A'})^N$ are such that $b(i)$ and $b'(i)$ agree on the $s$-th coordinate, then $(b, b') \in E_{\id_s}^{(\mathbf A')^N}$, and hence $(g(b), g(b')) \in E_{\id_s}^{\mathbf B'}$ which means that $g(b)$ and $g(b')$ agree on the $s$-th coordinate.
  This shows that $g$ decomposes as $\prod_{s\in \cat S} f_s$. The fact that the corresponding $f$ is a natural transformation $A^N \to B$ is straightforward to check.
\end{proof}

Given the above proposition, we can derive the Bulatov--Zhuk theorem for copresheaves using tools developed in Section~\ref{sec:bulatov-zhuk}.

\begin{proof}[Proof of Theorem~\ref{thm:bulatov-zhuk-for-presheaves}]
  If there is a morphism $\Ran_A A \to \id$, we get that $\hCSP(A)$ is \NP-hard by Corollary~\ref{cor:bulatov-zhuk-NP-complete}.
  Conversely, if there is no such morphism, let $\mathbf B = \mathbf A'$ be the structure constructed from the copresheaf $A$ as in Definition~\ref{def:single-sorted-from-presheaf}. We prove that $\hCSP(A)$ is polynomial-time solvable via a reduction to $\hCSP(\mathbf B)$.
  This reduction is obtained by Theorem~\ref{thm:ranGivesReductions} using the isomorphism $\pol(\mathbf B) \simeq \Ran_A A$ provided by Proposition~\ref{prop:presheaves-to-structures}.
  Formally, the theorem is only phrased for copresheaves, hence, we apply it to $\mathbf B^\vartriangle$ and $A$; note that $\Ran_{\mathbf B^\vartriangle} {\mathbf B^\vartriangle} \cong \pol(\mathbf B)$. We get a log-space reduction from $\hCSP(A)$ to $\hCSP({\mathbf B^\vartriangle})$. However, as in Section~\ref{sec:relational-structures}, $\hCSP({\mathbf B^\vartriangle})$ and $\hCSP(\mathbf B)$ are log-space equivalent, hence there is a log-space reduction from $\hCSP(A)$ to $\hCSP(\mathbf B)$.
  Furthermore, there is still no natural transformation $\pol(\mathbf B) \to \id$ since the former is isomorphic to $\Ran_A A$. Hence, Theorem~\ref{thm:bulatov-zhuk} provides that $\hCSP(\mathbf B)$, and hence also $\hCSP(A)$, is in~\Ptime.
\end{proof}

It might appear that there is possibly a different strategy to proving Theorem~\ref{thm:bulatov-zhuk-for-presheaves} since Theorem~\ref{thm:3.12} provides an equivalence between $\hCSP(A)$ and $\sPCSP(\id, M)$ for the monad $M = \Ran_A A$.
Hence, we could immediately derive the result if we knew that $\sPCSP(\id, M)$ is polynomial-time solvable for all monads $M\colon \fin \to \fin$ such that there is no natural transformation $M \to \id$.
Nevertheless, this statement is not a direct consequence of the Bulatov--Zhuk theorem, for example, since not all monads $M\colon \fin \to \fin$ are of the form $\pol(\mathbf B)$ for a finite relational structure $\mathbf B$. Also note that the reduction from $\sPCSP(\id, M)$ to $\hCSP(\mathbf B)$, which could provide tractability of the former problem, is efficient only on instances of bounded set sizes.
Despite that, many algorithms for $\sPCSP(\id, M)$ including algorithms based on linear programming, affine integer programming \cite[Section 7]{BBKO21}, and on their combination BLP+AIP \cite{BrakensiekGWZ20} are polynomial-time even without restriction on the sizes of the sets. This raises the following question.

\begin{question}
  Let $M \colon \fin \to \fin$ be a monad such that there is no natural transformation $M \to \id$. Is the problem $\sPCSP(\id, M)$ always in \Ptime?
\end{question}

Observe that if $M\colon \fin \to \fin$ is a monad which does not allow a natural transformation to $\id$, then $\sPCSP(\id, M)$ is polynomial-time solvable on instances of bounded set-size. This is because restricting both $\id$ and $M$ to a finite subcategory $\cat S$ of $\fin$ yields a tractable $\sPCSP(\id|_\cat S, M|_\cat S)$ by Theorem~\ref{thm:bulatov-zhuk-for-presheaves}.
Hence, an example of a monad $M$, without a natural transformation $M \to \id$ but with \NP-complete $\sPCSP(\id, M)$, would raise follow-up questions on the fixed-parameter tractability of the problem $\sPCSP(\id, M)$.
As far as we are aware, these questions have not been investigated before although the answers might be very relevant for the quest for the so-called \emph{uniform algorithm} for tractable CSPs.

\section{Conclusion}
  \label{sec:conclusion}

We have shown that translating combinatorial and algebraic notions of the CSP theory to the language of copresheaves leads to an elegant theory. Standard categorical constructions are a natural part of this theory and lead to simple and clean proofs of core statements of the algebraic approach to (P)CSPs.

We conclude the paper with an exposition of several directions for future research, which have substantial potential to bring new complexity results about CSPs and PCSPs.

\subparagraph{CSPs with functional symbols}
The study of CSPs for algebras has recently received some attention~\cite{BartoDM21, BartoM24}.
Due to a theorem of Pultr~\cite{Pul70}, Proposition~\ref{p:nerve-realization} applies to all locally finitely presentable categories, in particular to structures with both relational and algebraic signature. Therefore, results in Section~\ref{s:adjunctions} should generalise too.
It remains an open question whether Theorem~\ref{thm:bulatov-zhuk} applies to structures with operations of higher arity.

\subparagraph{Thin adjunctions characterise reductions}
It was shown in \cite{KOWZ23} that adjunctions between the posetification of the two categories can be used to characterise monotone reductions which preserve disjoint unions.
Hence an extension of our Section~\ref{sec:gadgets} to such adjunctions might provide new reductions beyond the gadget reductions. Some of such adjoint pairs were described in \cite{KOWZ23,DalmauKO24}.
Nevertheless, there are many questions that remain open: Which of the left adjoints are efficiently computable? Is there a characterisation of these more general reductions analogous to Theorem~\ref{thm:gadgets}?

\subparagraph{Topological methods in CSPs and PCSPs}
Our categorical methods are a natural setting for the use of topological methods in the theory of CSPs \cite{MeyerO25} and PCSPs \cite{KOWZ23,FilakovskyNOTW24, AvvakumovFOTW25}.
In fact, our framework immediately applies to (locally finite) simplicial sets, simplicial sets with an action of a group, etc., since all these are types of presheaves.
Templates that involve Kan complexes with finitely many non-trivial homotopy groups are of particular interest since a bound on the dimension of obstructions puts the problem in \NP.

\subparagraph{Presheaves of local solutions}
Abramsky \cite{Abramsky22} and Ó Conghaile~\cite{OConghaile22} provide a perspective on hierarchies of CSP algorithms through presheaves of local solutions.
Many hierarchical algorithms, e.g.\ local consistency, cohomological $k$-consistency, Sherali--Adams hierarchy and levels of the BLP+AIP algorithm, which have recently received a lot of attention, e.g.\ \cite{CiardoZ25-shadow,LichterP25,ChanNP24,ChanN25,ConnerydGP26}, can be then explained as an application of the local-solution presheaf construction followed by an algorithm solving $\sPCSP(\id, M)$ for a suitable $M\colon \fin \to \set$.
We believe that the categorical approach could provide new insights about the power of these and other similar hierarchies.

\appendix

\section{Logical formulation of CSPs}
\label{sec:translations}

There is a third possible formulation of a CSP which we have not discussed in the main part of the paper. Namely, viewing the CSP with template $\mathbf A$, which is a relational structure, as the problem of \emph{satisfaction of primitive positive (pp) sentences} in $\mathbf A$, that is, sentences in first-order logic built only using existentials, conjunctions and atoms.

\begin{definition}[PCSP, logic version]
  Fix a relational structures $\mathbf A$ and $\mathbf B$ such that there is a homomorphism $\mathbf A \to \mathbf B$, the problem $\ppPCSP(\mathbf A, \mathbf B)$ is a promise problem defined as follows.
  Given a pp-sentence $\phi$, output:
  \begin{description}
    \item[\Yes] if $\mathbf A \models \phi$,
    \item[\No] if $\mathbf B \not\models \phi$.
  \end{description}
\end{definition}

The equivalence of $\ppCSP(\mathbf A)$ and $\hCSP(\mathbf A)$ is given by the Chandra--Merlin correspondence~\cite{chandra1977optimal}.
Below we discuss insights that can be gained from this perspective, and the connections between our categorical formulations and the logical formulation above.

\begin{remark}
  The above definition makes sense for arbitrary first-order signatures (since homomorphisms preserve satisfaction of pp-sentences). Nevertheless, unlike for relational structures, it is not clear that $\ppCSP(\mathbf A)$ is equivalent to $\hCSP(\mathbf A)$ for structures with at least one at least binary function symbol.
\end{remark}

\subsection{Primitive positive interpretations}
  \label{sec:pp-interpretations}

Usually, the algebraic approach to CSP does not focus too much on the reduction itself, many of the results (e.g.\ \cite[Section 3.4]{BKW17} and \cite[Theorem 4.12 and Section~9]{BBKO21}) are phrased in terms of what happens to the templates, and use the notion of \emph{pp-interpretations} (or two related terms of \emph{pp-powers} and \emph{pp-constructions}) which we explain in this subsection together with their relations to nerves.

A \emph{primitive positive formula (pp-formula)} is a first-order formula constructed only using atoms, conjunctions, and existential quantifiers, e.g.\ the formula
\[
  (\exists u, v)\,
  E(x, u) \wedge E(u, v) \wedge R(x, v, y) \wedge x = y.
\]
We always write pp-formulae in prenex normal form.
A common alternative definition of a CSP whose template is a relational structure $\mathbf A$ is: Given a pp-sentence (i.e.\ pp-formula without free variables) $\phi$, decide whether $\mathbf A$ satisfies $\phi$; we use the symbol $\mathbf A \models \phi$ to denote that $\phi$ is satisfied in $\mathbf A$.
The equivalence of this formulation with the homomorphism formulation of Definition~\ref{def:csp-v1} is given by a correspondence between pp-formulae and relational structures, which is sometimes attributed to Chandra and Merlin \cite{chandra1977optimal}.
It may be formulated in several ways, we phrase it as a correspondence between quantifier-free pp-formulae (i.e.\ conjunctions of atoms) $\phi$ and structures $\mathbf C$ such that assignments to variables of $\phi$ which make the formula true in $\mathbf A$ are in bijection with homomorphisms $\mathbf C \to \mathbf A$.
The correspondence than assigns $\mathbf C$ to each conjunction of atoms $\phi$, and, conversely, a conjunction of atoms $\phi$ to each structure $\mathbf C$ such that (in either case) $\phi^\mathbf A \simeq \hom(\mathbf C, \mathbf A)$.

Logical interpretations are a standard way of defining a structure of some signature $\Pi$ from another structure of (usually) different signature $\Sigma$. In plain words, a structure $\mathbf B$ is said to be interpretable in $\mathbf A$ if the domain of $\mathbf B$ and all its relations are relations definable in $\mathbf A$. For pp-interpretations, we use the following notion of definability.
We write $\phi^\mathbf A$ where $\phi$ is a pp-formula with $k$ free variables for the relation $\phi^\mathbf A \subseteq (D^\mathbf A)^k$ defined by
\[
  \phi^\mathbf A(a_1, \dots, a_k)
  \text{ iff }
  \mathbf A, a_1, \dots, a_k \models \phi,
\]
i.e.\ $R$ consists of all tuples $(a_1, \dots, a_k)$ such that $\mathbf A$ satisfies $\phi$ after substituting $a_1, \dots, a_k$ for the free variables. We say that $R$ is \emph{pp-definable} in $\mathbf A$ if $R = \phi^\mathbf A$ for some pp-formula $\phi$.

\begin{definition}
  Let $\Sigma$ and $\Pi$ be relational signatures, an \emph{interpretation} $\phi$ from $\Sigma$-structures to $\Pi$-structures is given by formulae in the language of $\Sigma$-structures: $\phi_D$ (for the domain) with $n$ free variables, and a formula $\phi_R$ for each $k$-ary $\Pi$-symbol $R$ with $nk$ free variables such that
  \begin{equation} \label{eq:cons-int}
    \phi_R(x_1, x_2, \dots, x_{nk}) \to \phi_D(x_{in+1}, x_{in+2}, \dots, x_{(i+1)n})
  \end{equation}
  for all $i = 0, \dots, k-1$.

  Such an interpretation defines a functor $I_\phi$ from $\Sigma$-structures to $\Pi$-structures which maps a $\Sigma$-structure $\mathbf A$ to a $\Pi$-structure $I_\phi(\mathbf A)$ defined as:
  \begin{align*}
    D^{I_\phi(\mathbf A)} &= \phi_D^\mathbf A \\
    R^{I_\phi(\mathbf A)} &= \phi_R^\mathbf A &\text{for each $\Pi$-symbol $R$}
  \end{align*}
  where $\phi_R^\mathbf A$ is interpreted as a relation on $\phi_D^\mathbf A$ of arity $k$ in the natural way, i.e.\ it consists of all tuples
  \[
    ((a_1, \dots, a_n), (a_{n+1}, \dots, a_{2n}), \dots, (a_{(k-1)n+1}, \dots, a_{kn}))
  \]
  such that $(a_1, \dots, a_{kn}) \in \phi_R^\mathbf A$.
  
  An interpretation $\phi$ is a \emph{pp-interpretation} if all formulae $\phi_D$ and $\phi_R$ are primitive positive.
  We say that a structure $\mathbf B$ is \emph{pp-interpretable} from $\mathbf A$ if there exists a pp-interpretation $\phi$ such that $\mathbf B$ is isomorphic to $I_\phi(\mathbf A)$.
\end{definition}

\begin{example} \label{ex:c5-k5}
  We claim that the graph $K_5$ is pp-interpretable from $C_5$. The pp-interpretation $\phi$ witnessing this fact is given by formulae:
  \begin{align*}
    \phi_D(x)    &\text{ iff } x = x\\
    \phi_E(x, y) &\text{ iff }
      (\exists u, v)\; E(x, u) \wedge E(u, v) \wedge E(v, y)
  \end{align*}
  Observe that $I_\phi$ does not change the domain of the structure, in particular the domain of $I_\phi(C_5)^\triangledown$ is the same as the domain of $C_5$. Two vertices $u, v$ are then connected with an edge in $I_\phi(C_5)^\triangledown$ if there is a path of length $3$ in $C_5$ that connects $u$ and $v$. It is easy to observe that this is indeed true for any $u \neq v$, and hence $I_\phi(C_5)^\triangledown$ is a 5-clique.
\end{example}

Observe that, in the example above, the fact that $(u, v) \in I_\phi(C_5)$ can be witnessed in several ways by choosing different assignments to the free variables $z_1, z_2$. In that sense, it is more natural to define the resulting graph as a multigraph.
In general, with a few additional syntactic restrictions on the formulae appearing in a pp-interpretation, we may define the interpretation functor to be valued in $[\cat R_\Pi, \fin]$.
These additional restrictions change the expressive power of pp-interpretations only slightly; it can be shown that for each pp-interpretation $\phi$, there is a pp-interpretation $\phi'$ of the required form such that $I_\phi(\mathbf A)$ and $I_{\phi'}(\mathbf A)$ are homomorphically equivalent for all $\mathbf A$.

\begin{definition}
  Let $\phi$ be an interpretation from $\Sigma$-structures to $\Pi$-structures, and assume that moreover that
  \begin{itemize}
    \item $\phi_D$ is quantifier free; and
    \item $\phi_R$ is of the form $(\exists z_1, \dots, z_k)\, \phi_R^\text{qf}$ where $\phi_R^\text{qf}$ is a conjunction of atoms.
  \end{itemize}
  We then define a copresheaf $B\colon \cat R_\Pi \to \fin$ by $B(D) = \phi_D^\mathbf A$  and $B(R) = (\phi_R^\text{qf})^\mathbf A$, where $B$ on the maps is defined in the natural way, i.e.\ $B(p_R^i)\colon (\phi_R^\text{qf})^{\mathbf A} \to \phi_D^{\mathbf A}$ is the projection to the variables $(x_{id+1}, \dots, x_{(i+1)d})$.
  We denote the resulting copresheaf $B$ by $I^\vartriangle_\phi(\mathbf A)$.
\end{definition}

Note that the only real assumption above is that $\phi_D$ is quantifier free since any pp-formula $\phi_R$ can be rewritten to an equivalent formula of the required form.
Observe that $(I^\vartriangle_\phi(\mathbf A))^\triangledown = I_\phi(\mathbf A)$, i.e.\ retyping the result as a $\Pi$-structure recovers $I_\phi$ from $I^\vartriangle_\phi$.

\begin{proposition}
  Let $\Sigma$ and $\Pi$ be relational signatures.
  \begin{enumerate}
    \item For each pp-interpretation $\phi$ from $\Sigma$-structures to $\Pi$-structures such that $\phi_D$ is quantifier free, there exists $G\colon \cat R_\Pi \to \str\Sigma$ such that, for all $\Sigma$-structures $\mathbf A$,
      \begin{equation} \label{eq:pp-nerve}
        \nerve G(\mathbf A) \simeq I_\phi^\vartriangle(\mathbf A).
      \end{equation}
    \item For each functor $G\colon \cat S \to \str\Sigma$, there exists a pp-interpretation $\phi$ from $\Sigma$-structures to $\Pi$-structures satisfying \eqref{eq:pp-nerve}.
  \end{enumerate}
\end{proposition}

\begin{proof}[Proof sketch]
  \begin{enumerate}
    \item
      Given the pp-interpretation $\phi$, we define $G$ using the Chandra--Merlin correspondence.

      We first define the structure $G(D)$ from the formula $\phi_D$. Note that $\phi_D$ is a conjunction of atoms.
      For simplicity, consider the case that $\phi_D$ has no atoms of the form $x_i = x_j$.
      Then $G(D)$ is defined as the $\Pi$-structure whose domain $G(D)$ is the set of all variables in $\phi_D$, and the relation $R^{G(D)}$ contains all tuples $(x_1, \dots, x_k)$ such that $\phi_D$ contains the atom $R(x_1, \dots, x_k)$. It is not hard to check that satisfying assignments to $\phi_D$ in a structure $\mathbf A$, i.e.\ elements of $\phi_D^\mathbf A$, are in 1-to-1 correspondence with homomorphisms $G(D) \to \mathbf A$.
      If $\phi_D$ includes equality, $G(D)$ is constructed similarly with the exception that if $\phi_D$ contains $x = y$, then the elements corresponding to $x$ and $y$ are identified. In other words, $G(D)$ is the factor of the above structure by the equivalence relation defined by the equality atoms in $\phi_D$.

      For a $\Sigma$-symbol $R$, $G(R)$ is defined in the same way except we use the formula $\phi_R^\text{qf}$.
      Furthermore, the morphism $G(p_i^R) \colon G(D) \to G(R)$ maps the elements of $G(D)$ corresponding to variables $x_1, \dots, x_n$ to the elements of $G(R)$ corresponding to variables $x_{in+1}, \dots, x_{(i+1)n}$, respectively. This mapping preserves the relations (including the forced equality) because of \eqref{eq:cons-int}.

      The isomorphism between $\nerve G(\mathbf A)$ and $I_\phi(\mathbf A)$ is given by the bijections $\hom(G(X), \mathbf A) \simeq \phi_X^\mathbf A$. Checking that this indeed gives a natural equivalence is straightforward.
    \item
      Given a functor $G\colon \cat R_\Pi^\op \to \str\Sigma$, we construct the pp-interpretation $\phi$ by reversing the process above.
      For simplicity assume that the domain of $G(D)$ is $\{1, \dots, n\}$.
      The define $\phi_D(x_1, \dots, x_n)$ as the conjunction of atoms $S(x_{a_1}, \dots, x_{a_m})$ for which we have $S^{G(D)}(a_1, \dots, a_m)$.
      Similarly, we might want to define $\phi_R$ as the conjunction of atoms of the form $S(y_{b_1}, \dots, y_{b_m})$, but we also need to make sure that the free variables of $\phi_R$ align with $k$-copies of free variables of $\phi_D$ (assuming $R$ is of arity $k$).
      Instead, we define $\phi_R$ as
      \[
        \phi_R(x_1, \dots, x_{nk}) \text{ iff }
          (\exists y_b, \dots)\,
          \phi'_R \wedge
          \bigwedge_{\substack{i = 1, \dots, k\\ d = 1, \dots, n}} x_{(i-1)n+d} = y_{G(p_i^R)(d)}
      \]
      The existentially quantified variables $y_b$ range through all elements $b$ of $G(R)$,
      and $\phi'_R$ is the conjunction of atoms $S(y_{b_1}, \dots, y_{b_k})$ such that $S^{G(R)}(b_1, \dots, b_k)$.
      We leave out checking that the formulae satisfy \eqref{eq:cons-int}, and that the resulting nerve satisfies \eqref{eq:pp-nerve}.
      \qedhere
  \end{enumerate}
\end{proof}

\begin{example}[Example~\ref{ex:c5-k5} cont.]
  Let us describe construction of $G$ from the interpretation given in Example~\ref{ex:c5-k5}.
  The graph $G(D)$ is simple, it consist of a single vertex $x$ and no edges as $\phi_D$ does not have other variables, and no atoms of the form $E(x, y)$.
  The graph $G(E)$ is then the following graph:
  \[
    \begin{tikzpicture}
      \node (0) {$x$};
      \node (1) at (1, 0) {$z_1$};
      \node (2) at (2, 0) {$z_2$};
      \node (3) at (3, 0) {$y$};
      \draw (0) -- (1); \draw (1) -- (2); \draw (2) -- (3);
    \end{tikzpicture}
  \]
  The mapping $G(p_1^E) \colon G(D) \to G(R)$ maps $x$ to $x$, and the mapping $G(p_2^E) \colon G(D) \to G(R)$ maps $x$ to $y$.
  It is not hard to check that indeed $N_G(A) \simeq I_\phi^\vartriangle(A)$ for all graphs $A$.
\end{example}

The above proposition has an important corollary, namely that any right adjoint between categories of relational structures, or more precisely from relational structures to copresheaves corresponding to relational structures (of a possibly different signature) is given by a pp-interpretation.
This is an immediate consequence of Proposition~\ref{p:nerve-realization}.

\begin{corollary}
  For every quantifier-free pp-interpretation $\phi$ from $\Sigma$-structures to $[\cat R_\Pi, \fin]$, $I_\phi^\vartriangle$ admits a left adjoint of the form~$\kay G$.

  Moreover, any functor $R \colon \str\Sigma \to [\cat R_\Pi, \fin]$ which admits a left adjoint is naturally equivalent to $I^\vartriangle_\phi$ for some pp-interpretation $\phi$.
\end{corollary}

An explicit description of the left adjoint for the case $\cat S = \cat R_\Pi$ for some relational signature can be found in \cite[Theorems 13 and~15]{BKW17}.
For example, the `left adjoint' $L$ to $I_\phi$ where $\phi$ is the interpretation from Example~\ref{ex:c5-k5} is given by subdividing each edge of a given graph in three. Note that between categories of graphs, it is adjoint only in the weaker sense: $L(G) \to H$ if and only if $G \to I_\phi(H)$, but the correspondence is not 1-to-1.

\begin{remark}
  Several other notions closely related to pp-inter\-pre\-ta\-tions are often used in literature on complexity of (P)CSPs, e.g.~\cite{BartoOP18,BKW17,BBKO21}:
  \emph{pp-powers} are pp-interpretations where $\phi_D$ includes no relational symbol, and hence $\phi_D^\mathbf A = (D^\mathbf A)^n$ for some $n$.
  A structure $\mathbf A$ is said to be \emph{pp-constructible} from $\mathbf B$ if $\mathbf A$ is homomorphically equivalent to a pp-power of $\mathbf B$, and a promise template $\mathbf A, \mathbf A'$ is \emph{pp-constructible} from another promise template $\mathbf B, \mathbf B'$ if there is a~pp-interpretation $\phi$ with $\mathbf A \to I_\phi(\mathbf B)$ and $I_\phi(\mathbf B') \to \mathbf A'$.
\end{remark}

\bibliographystyle{plainurl}

\end{document}